\documentclass[aps,superscriptaddress,pra,twocolumn,footinbib]{revtex4-1}
\usepackage{amsmath,amssymb}
\usepackage{bm}
\usepackage{epsfig}
\usepackage{natbib}
\usepackage{hyperref}

\usepackage{graphicx}
\usepackage{epstopdf}
\usepackage{color,xcolor}
\usepackage{bm}
 \hypersetup{pdfstartview={FitH},pdfpagemode={UseNone},
            colorlinks,linkcolor=red, citecolor=blue, urlcolor=blue,
            bookmarks=true, bookmarksopen=true, pdfnewwindow=true}
             
\usepackage{verbatim} 
\usepackage{morefloats}
\usepackage{bibentry}
\usepackage[mathscr]{euscript}

\renewcommand{\phi}{\varphi}

\begin{document}
\title{Boosted second-harmonic generation in the LiNbO\textsubscript3 metasurface governed by high-Q guided resonances and bound states in the continuum}
\author{Ze Zheng}
\affiliation{Advanced Optics and Photonics Laboratory, Department of Engineering, School of Science \& Technology, Nottingham Trent University, Nottingham NG11 8NS, UK}%
\author{Lei Xu}%
\affiliation{Advanced Optics and Photonics Laboratory, Department of Engineering, School of Science \& Technology, Nottingham Trent University, Nottingham NG11 8NS, UK}%
\email{lei.xu@ntu.ac.uk}

\author{Lujun Huang}
\affiliation{School of Engineering and Information Technology, University of New South Wales, Canberra ACT 2600, Australia}%
\author{Daria Smirnova}
\affiliation{ARC Centre of Excellence for Transformative Meta-Optical Systems (TMOS), Research School of Physics, The Australian National University, Canberra, ACT 2601, Australia}%
\author{Peilong Hong}
\affiliation{School of Optoelectronic Science and Engineering, University of Electronic Science and Technology of China (UESTC), Chengdu 611731, China}%
\altaffiliation[Also at ]{The Key Laboratory of Weak Light Nonlinear Photonics, Nankai University, Tianjin 300457, Ministry of Education, China}%
\author{Cuifeng Ying}%
\affiliation{Advanced Optics and Photonics Laboratory, Department of Engineering, School of Science \& Technology, Nottingham Trent University, Nottingham NG11 8NS, UK}%
\author{Mohsen Rahmani}%
\affiliation{Advanced Optics and Photonics Laboratory, Department of Engineering, School of Science \& Technology, Nottingham Trent University, Nottingham NG11 8NS, UK}%

\begin{abstract}
To date, second-harmonic generation (SHG) at nanoscale has been concentrated on employing high-refractive-index nanostructures, owing to the strong field confinement at deep subwavelength scales based on optically resonant effects. However, low-index nanostructures generally exhibit weaker resonant effects and lower field confinement. To address this issue, by harnessing the large nonlinearity of LiNbO\textsubscript3, we propose a novel approach to employ guided resonances and bound states in the continuum (BICs) with a LiNbO\textsubscript3 metasurface consisting of a LiNbO\textsubscript3 disk array sitting on a LiNbO\textsubscript3 thin film. Such a system can transform the guided modes supported by LiNbO\textsubscript3 thin film into high-quality guided resonances which can be excited directly under plane-wave illumination. Importantly, we further demonstrate strong field confinement inside LiNbO\textsubscript3 thin film with tailorable Q-factor by realising a Friedrich-Wintgen BIC. Such a unique mode engineering enables a record-high SHG efficiency of 5\% under a pump intensity as low as 0.4 $\mathrm{MW/cm^{2}}$. Moreover, we reveal the influence of nonlinear resonances and cross-coupling on the SHG by showing the anomalous SHG and efficiency tuning with the rotation of the crystal axis. Our work offers a new route to constructing enhanced SHG based on high-Q guided resonances and BICs, including low-index and high-index nonlinear materials.
\end{abstract}

\maketitle


\section{\label{sec:level1}Introduction}

Frequency conversion is one of the most intensively studied areas in nonlinear optics at the macro-scale \cite{boyd2020nonlinear} due to its extensive range of applications, including but not limited to sensing \cite{mesch2016nonlinear,butet2012sensing}, quantum light source \cite{moody2020chip}, and nonlinear imaging \cite{schlickriede2020nonlinear,burns2000nonlinear}. High-power lasers and bulk nonlinear crystals have been widely used in nonlinear optics in the last decades to generate strong nonlinear signals. This technology exploits phase-matching to maximise the nonlinear conversion efficiency. However, nonlinear optics at the nanoscale has not been fully developed yet due to the minor mode volume for light-matter interactions and, subsequently, the low conversion efficiency. In the past decades, the development of technology for the growth and nano-fabrication of nonlinear material has enabled various applications, one of which is the resonance-enhanced harmonic generation \cite{vabishchevich2018enhanced,kruk2015enhanced}. The novel conception, utilising the strongly resonant modes built in the light-matter in the photonic nanostructure to enhance the nonlinear generation, has been widely investigated and developed in recent years \cite{zheludev2012metamaterials,kauranen2012nonlinear,rahmani2018nonlinear}. Importantly, phase-matching is no longer necessary in the resonant structures due to their subwavelength size \cite{carletti2019high}. 

Metallic nanoparticles, working based on collective oscillations of the free electrons, so-called surface-plasmons, have been intensively studied for nonlinear interactions at the nanoscale in the last decade~\cite{kauranen2012nonlinear}. However, the high Ohmic loss possessed by metals limits their practical use \cite{kuznetsov2016optically}. Consequently, all-dielectric nanoparticles with high refractive indices, large nonlinear coefficients, and low losses have been used in building enhanced optical nonlinear processes over the past few years. Such nanoparticles support Mie resonances which enable the enhanced harmonic generations, including second-harmonic generation (SHG), third-harmonic generation (THG), and four-wave mixing (FWM) \cite{bonacina2020harmonic,carletti2015enhanced,grinblat2021nonlinear,shcherbakov2014enhanced,grinblat2016enhanced,locher2018polarization,liu2016resonantly,YudongLu2022Opto,gigli2022all}. Mie-type resonances can provide strong local field confinement inside the nanostructure and thus enhance the nonlinear conversion efficiency from the resonant nanostructures \cite{doi:10.1021/acsphotonics.7b01277,frizyuk2019second}. Typically, the quality-factor (Q-factor) of Mie resonances becomes the principal index to indicate the capability of local electromagnetic field confinement.

The Q-factor of resonance is defined as energy dissipation each cycle of oscillation versus the energy stored in the resonator. A higher Q-factor is expected to induce stronger field confinement within the resonator, leading to larger light-matter interactions \cite{liang2020bound,zeng2021light}. Recently, bound states in the continuum (BICs) have attracted tremendous attention for their ability to achieve resonances with the infinite Q-factors that can significantly increase the nonlinear conversion efficiency \cite{carletti2019high,koshelev2020subwavelength,yang2020nonlinear}. Relevant works are demonstrated and proposed in nonlinear optical processes based on different high-index materials, including silicon, GaAs, GaP, etc \cite{vabishchevich2018enhanced,carletti2019high,tong2016enhanced,anthur2020continuous,koshelev2019nonlinear,xu2019dynamic}. However, most of the above dielectric and semiconductor materials are lossy, i.e. non-transparent, in the visible spectral range or the near-ultraviolet spectrum region, leading to the degradation in the performance for nonlinear applications. 

Recently, low-index nonlinear materials, such as AlN, GaSe and LiNbO\textsubscript3 that have large second-order nonlinear susceptibilities, have attracted increasing attention as they are highly transparent from near-infrared to near-ultraviolet spectrum region \cite{fujii1977nonlinear,allakhverdiev2009effective,jiang2020high,carletti2019second,vakulov2020oxygen}. Over the past few years, the progress of exploiting Mie resonance-based nanostructures and metasurfaces has enabled researchers to achieve enhanced nonlinear generations in these materials. The field confinement of Mie resonances in the low-index nanoparticles ($n \sim 2.0$ ) is usually lower than that in the high-index nanoparticles ($n > 3$), leading to an inefficient conversion for low-index nonlinear materials, like AlN, GaSe, and LiNbO\textsubscript3 \cite{kuznetsov2016optically}. 

In addition to nanoparticles, BIC states can also be formed by guided modes steadily existing in a slab with subwavelength thickness~\cite{gao2016formation}. The optical BIC states are observed experimentally both in silicon ($n = 3.48$) \cite{yin2020observation} and silicon nitride ($n = 2.02$) \cite{hsu2013observation}, indicating that both low and high-index materials can support the BIC states with divergent Q-factor. Besides, the quasi-BIC states and the guided modes can be easily excited by tuning the incident angle without any geometric adjustment. Compared to nanoparticles, the slab provides a larger volume of confined light fields to enhance the light-matter interactions.

Here, we propose a novel approach employing the best of both worlds: nanoparticles and slabs. We demonstrate that by accommodating a LiNbO\textsubscript3 disk metasurface on the top of a LiNbO\textsubscript3 thin film, the system can transform the guided modes supported by LiNbO\textsubscript3 thin film into high-Q guided resonances with strong field confinement via plan-wave illumination. Furthermore, we found that such a metasurface can host a Friedrich-Wintgen BIC formed by the interference between high-Q transverse-electric (TE) and transverse-magnetic (TM) odd modes. We reveal the formation mechanisms of these guided resonances and BICs and optimise the second-harmonic generation by controlling three processes: the field confinement at pump frequency, the resonance at harmonic frequency, and the coupling between the pump and harmonic waves.  Within the three processes, we demonstrate the tailorable Q-factor of resonance by tuning the incident angle and subsequently obtained highly boosted SHG emission. Our approach can be applied to a wide range of nonlinear materials with various nonlinear processes, which can expand a range of nonlinear photonics applications.

\section{Results}
\subsection{Design and approach}

\begin{figure*}[h]
\centering
\includegraphics[width=0.8\textwidth]{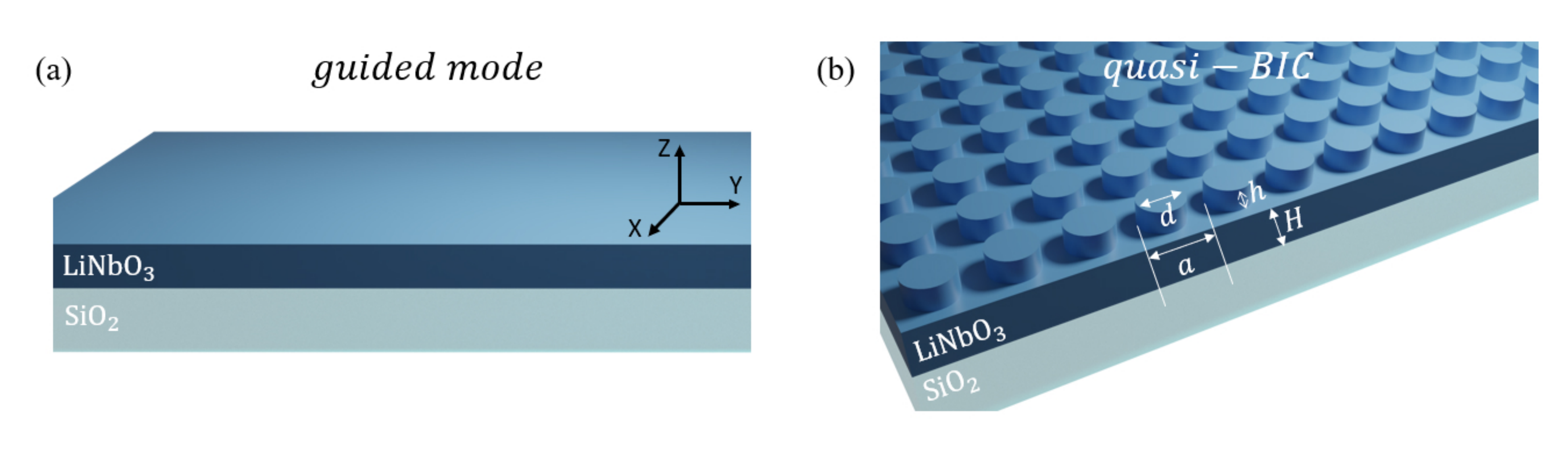}
\caption{\label{fig1}The schematics of LiNbO\textsubscript3 (a) slab and (b) metasurface-slab system, representing the guided mode and quasi-BIC mode.}
\end{figure*}

Our design consists of an infinite LiNbO\textsubscript3 slab on a silicon dioxide substrate as shown in Fig. \ref{fig1}(a). For simplicity, here the refractive indexes of LiNbO\textsubscript3, silicon dioxide, and air are set as $n_{slab}=2.2264$ \cite{zelmon1997infrared}, $n_{sub}=1.5$, and $n_{air}=1$, respectively. The slab and the substrate are uniform along the x-axis and y-axis. The system is invariant by a translation through any displacement $\mathbf{d}$ in the xy plane. We define the operator $\mathbf{T_d}$ as the displacement translation, therefore, the permittivity distribution is invariant under the displacement translation as shown in the equation: $\mathbf{T_d}\varepsilon(\mathbf{r})=\varepsilon(\mathbf{r}+\mathbf{d})=\varepsilon(\mathbf{r})$. The field amplitudes of the radiative modes are the same in the xy plane due to the continuous translational symmetry (Supplementary Material S1) \cite{joannopoulos2011photonic}. Following this rule, we can deduce the form for the radiative modes: $\mathbf{F_k}(\mathbf{r})=e^{i\mathbf{k}\boldsymbol{\rho}}\mathbf{f}(z)$, where $\mathbf{k}$ is the wave vector and $\mathbf{\rho}$ is the vector that is confined to the xy plane: $\boldsymbol{\rho}=x\mathbf{i}+y\mathbf{j}$. Since the refractive index of slab material (LiNbO\textsubscript3) is higher than its surroundings (air and silicon dioxide), there are an infinite number of guided modes that enables the oscillation of photons inside the slab. Those guided modes are decoupled to the modes that radiate to the far field with the field $\mathbf{F_k}(\mathbf{r})=e^{i\mathbf{k}\boldsymbol{\rho}}\mathbf{f}(z)$ due to the continuous translational symmetry in x and y directions.

\begin{figure*}
\centering
\includegraphics[width=0.85\textwidth]{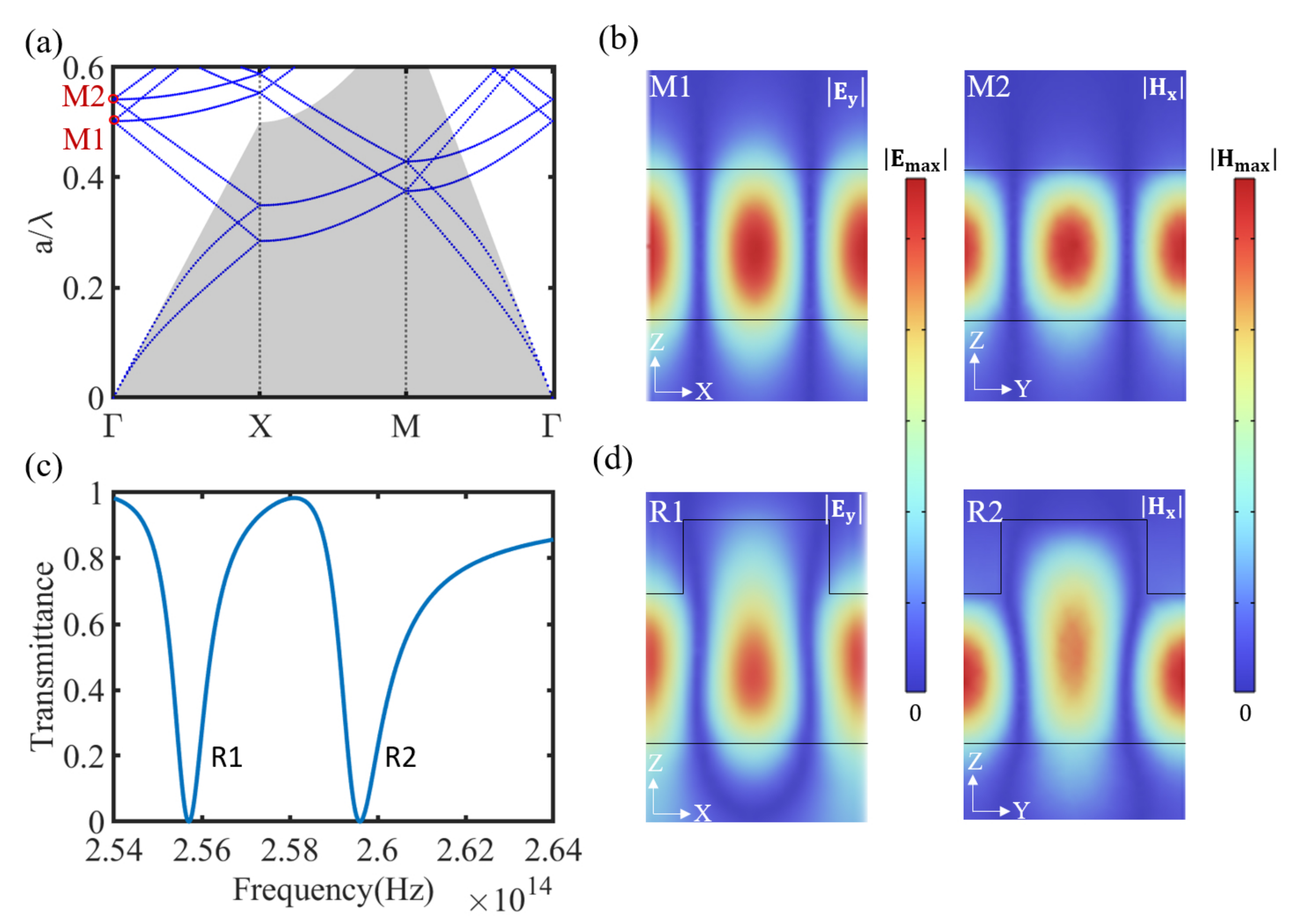}
\caption{\label{fig2}(a) The calculated band structure of the infinite slab. (b) The electric patterns of the guided modes M1 and M2 within the infinite slab at gamma point. (c) The transmittance spectrum of metasurface-slab at gamma point. The height of the slab is 400 nm. The height of the disk is 200 nm. The period length is 600 nm. The radius of the disk is 200 nm. (d) The electric patterns of the excited resonances R1 and R2 within the metasurface-slab.}
\end{figure*}

The dispersion relation of these guided modes can be obtained by calculating the band structure. It is worth noting that by using different periodicity, the dispersion relation will be the same as they give the same unfolded band structures. Here, we fix the periodicity as $a=600~\mathrm{nm}$, and then calculated the band structure from the slab structure. The periodic distribution has the form: $\mathbf{F}(\mathbf{r})=\mathbf{F}(\mathbf{r}+a\mathbf{i}+a\mathbf{j})$, where $\mathbf{i}$ and $\mathbf{j}$ represent the unit vector along the x- and y-axis. Fig. \ref{fig2}(a) indicates the band structure calculated under the periodic boundary condition. Here, we focus on two guided modes: one TE mode (M1) and one TM mode (M2) with field distributions shown in Fig. \ref{fig2}(b) and Supplementary material S2. The spectral position of these two modes can be roughly predicted as \cite{huang2013general,huang2021pushing}
\begin{eqnarray}
\frac{n\omega{a_x}}{c}\sim(l-1)2\pi+(m-1)2\pi\frac{a_x}{a_y},
\label{eq.1}
\end{eqnarray}
where $l$ and $m$ correspond to the antinode number along the x-axis and y-axis of the electric field in TE mode, and the magnetic field in TM mode. $a_x$ and $a_y$ correspond to the pitch size along x and y directions. In this study, we focus on the influence of variety of modes along x and y-axis rather than z-axis. Hence, we quoted the 2-D model (Eq.~\ref{eq.1}) instead of 3-D model to demonstrate and predict the spectrum positions of M1 and M2. For each propagating period, the guided modes require the one phase shift to be an integer multiple of $2\pi$. Here we set $a_x=a_y=a$. Eq.~\ref{eq.1} predicts that M1 (TE (l=3,m=1)) and M2 (TM (l=1,m=3)) are both located at the spectral position where $a/\lambda=1/n_{slab}$ , as indicated in Fig. \ref{fig2}(a). This provides a robust approach to access the guided modes at any wavelength by introducing and tuning this artificial periodicity $a$.  The introduction of the periodic nanodisks on top will break the symmetry of the system and then transform the guided modes M1, M2 into high-Q guided resonances which are accessible externally by plan-wave illumination. 

 The nanoparticle array on top enable the excitation of the M1 and M2 by leading more radiative modes into the system. Those radiative modes become the leaky channels between the guided modes and far-field, resulting in the high-Q guided resonances with tunable Q-factors. To excite the guided modes M1 and M2, we introduce metasurface-slab made of LiNbO\textsubscript3 on top of the infinite silicon dioxide substrate, as shown in Fig. \ref{fig1}(b). We respectively set the geometric parameters as: the height of the slab $H=400~\mathrm{nm}$, the height of disk $h=200~\mathrm{nm}$, the radius of disk $r=200~\mathrm{nm}$ and the periodic length $a=600~\mathrm{nm}$. The whole structure can be considered as adding a LiNbO\textsubscript3 disk on each periodic unit, $a\times a$, to transform the continuous transnational symmetry to discrete transnational symmetry (from $\mathbf{T_d}$ into $\mathbf{T}_a$). LiNbO\textsubscript3 is an anisotropy medium with the refractive index of $n_o=2.2264$ along the ordinary axis, and $n_e=2.1506$ along the extraordinary axis \cite{zelmon1997infrared}. Here we align the extraordinary axis of the LiNbO\textsubscript3 to the y-axis, so that $n_o=n_x=n_z=2.2264$, and $n_e=n_y=2.1506$. The system of the metasurface-slab is invariant under the $180^\circ$ rotation around the z-axis, which can be defined as an operator $C_2^z \varepsilon(x,y,z)=\varepsilon(-x,-y,z)=\varepsilon(x,y,z)$. 
 
 \begin{figure*}
\centering
\includegraphics[width=0.85\textwidth]{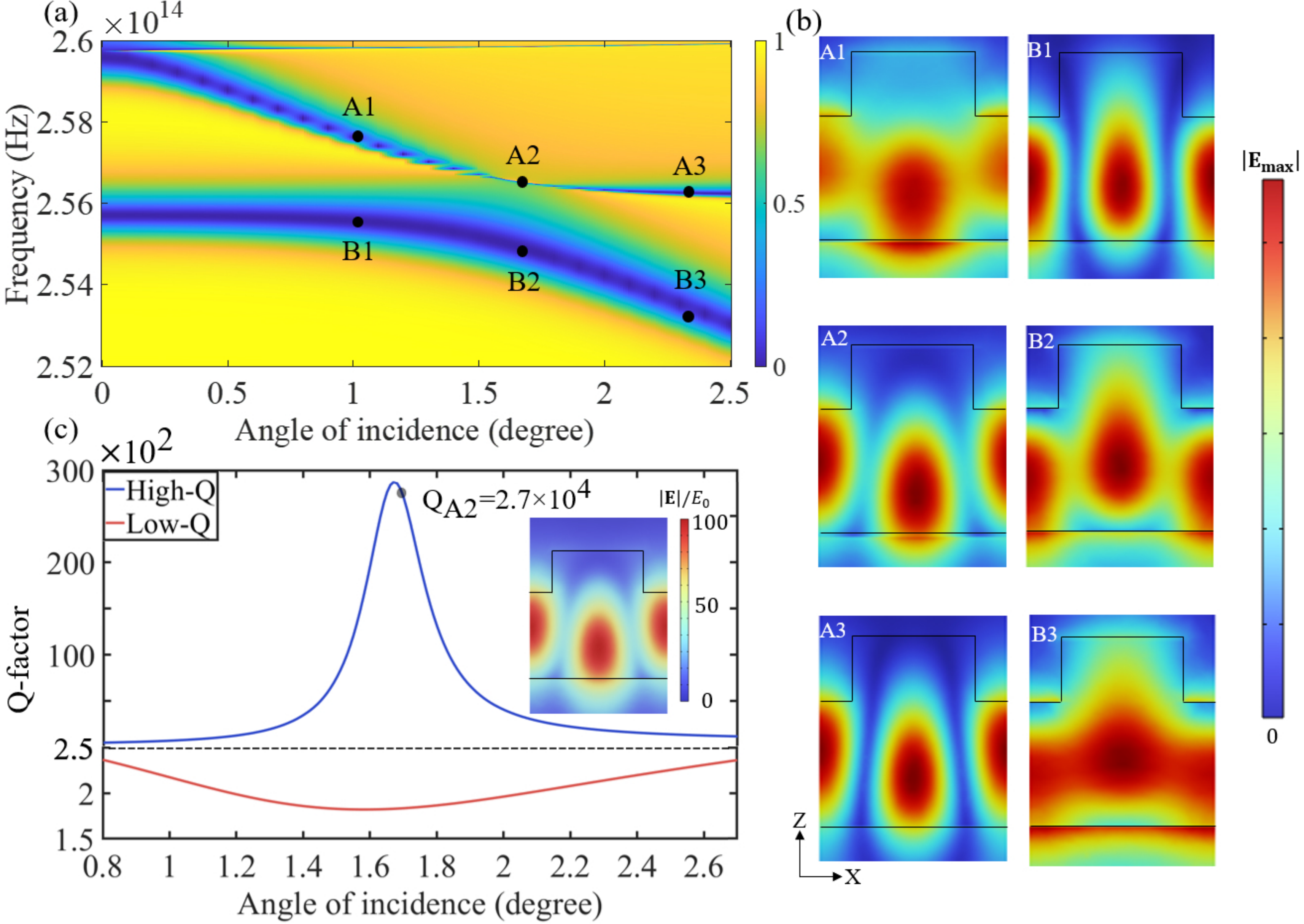}
\caption{\label{fig3}The transmittance spectrum of metasurface-slab with respect to incident angle (yz-plane). The polarisation of incident light is along the y direction. (b) The electric patterns of the high-Q resonance (R2) and low-Q resonance (R1) in different incident angles. (c) Q factors of coupled resonances R1 and R2 versus incident angle with the electric field distribution at 1.7-degree incident angle. }
\end{figure*}

 At the centre of the momentum space $((k_x,k_y)=(0,0))$, $\Gamma$, the field distributions of radiative channels along the z-axis are odd under the $C_2^z$  symmetry, and can couple to the odd guided mode within the slab and excite the odd resonances with finite Q-factor. The even guided modes are symmetry-protected BICs under $C_2^z$. They can be transformed into quasi-BICs by changing the incident angle with non-zero $k_x$ or $k_y$. The field distributions of Resonance 1 and Resonance 2 (Fig. \ref{fig2}(d) and Supplementary material S2) share similar field patterns of M1 and M2, indicating that R1 and R2 (shown in Figs. \ref{fig2}(c)(d)) excited under the y-polarized plane wave are the guided mode resonances transformed from the TE odd guided mode (M1) and TM odd guided mode (M2) respectively. In this way, we excited the odd guided resonances with high Q-factor and tailorable spectral position by engineering the periodicity of our system.  The process of designing a resonance from a continuous finite system to the periodic structure is always neglected in current research, but this process is also robust for obtaining desired light-matter interaction with both low- and high-index materials to achieve various applications.

The R1 and R2 possess Q-factors of several hundreds, having strong confinement of the electromagnetic field inside the metasurfaces. However, to significantly boost the light-matter interactions from the low-index-material-based nanostructures and metasurfaces, we further improve the confinement of the electromagnetic field of R1 by constructing the Friedrich-Wintgen BIC configuration based on R1 and R2 resonances. According to coupled mode theory, two resonances (here R1 and R2) coupled to the same radiative channels can strongly interfere with each other when they are close both spatially and spectrally. Under certain conditions (Supplementary material S3), one of them can turn into the BIC state with divergent Q-factor and vanished resonant linewidth when the radiation channels are completely suppressed via the destructive interference between the two resonances. The formation of this kind of BIC and quasi-BIC can be explained by a two-level system \cite{wiersig2006formation}.

According to Eq.~\ref{eq.1}, the spectral position of R1 and R2 can be well controlled by changing the periodic length $a$. However, tuning geometric parameters (also shown in Supplementary material S4) requires repeating the fabrication process to obtain the resonance at the ideal frequency. Instead, changing the incident angle can accomplish the same goal of controlling the spectral positions of the resonances by engineering the effective optical path length. We define the incident angle as the angle between the wave vector k and the normal to the plane of metasurfaces. We denote the incident angle as ‘the incident angle (xz-plane)’ when the wave vector is within the xoz-plane and ‘the incident angle (yz-plane)’ when the wave vector is within the yoz-plane. 

\begin{figure*}
\centering
\includegraphics[width=0.85\textwidth]{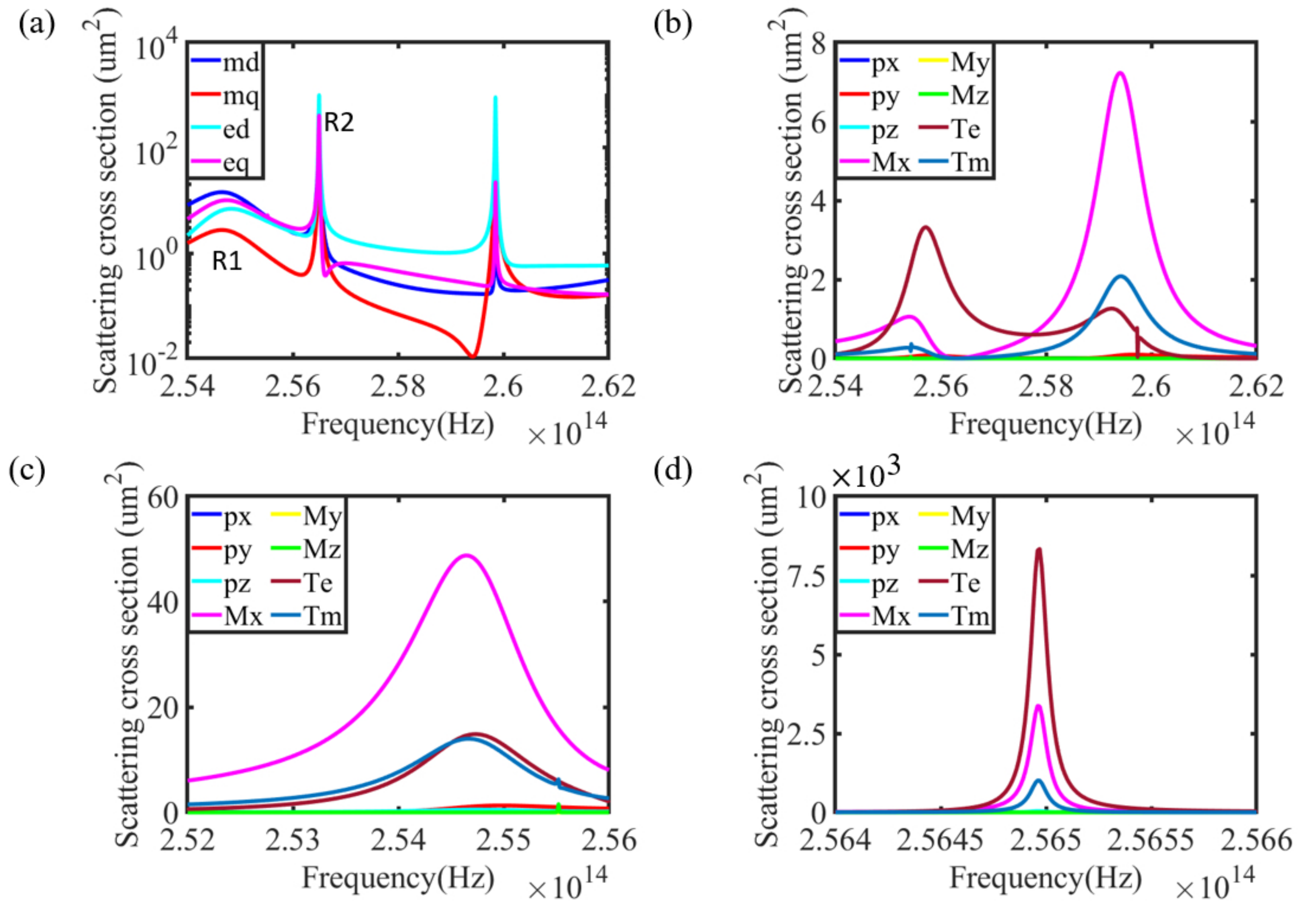}
\caption{\label{fig4}(a) The scattering cross section of electric dipole (ed), magnetic dipole (md), electric (eq) and magnetic quadrupole (mq) based on spherical multipole expansion under the 1.7-degree incident angle with y-polarized light. (b) The scattering cross section of R1 and R2 at the gamma point based on Cartesian multipole expansion including electric dipole in x-direction (px), y-direction (py), and z-direction (pz); magnetic dipole in x-direction (Mx), y-direction (My), and z-direction (Mz); electric (Te) and magnetic Toroidal moment (Tm). (c)(d) The scattering cross section of R1 (left) and R2 (right) under 1.7-degree incident angle with y-polarized light based on Cartesian multipole expansion.}
\end{figure*}

The spectral positions of the resonances are closely related to the incident angle and the electric field distributions of the mode. Based on Eq.~\ref{eq.1}, the spectral position of R1 is heavily dependent on structural size $a_x$ where the phase shift of an integer multiple of $2\pi$ is accumulated at the two interfaces, while the spectral position of R2 is highly dependent on structural size $a_y$. Thus, when the incident angle (yz-plane) increases, the frequency of R2(l=1,m=3) decreases, while R1(l=3,m=1) remains at the same frequency as shown in the transmittance spectrum under y-polarized light in Fig. \ref{fig3}(a). In this way, the spectral gap between the two resonances can be well controlled. The same control can be achieved under the x-polarized plane wave. When the incident angle (xz-plane) increases, the frequency of the resonance R1 decreases, while the frequency of the resonance R2 remains nearly unchanged (Supplementary material S5). By tuning the incident angle (yz-plane), R1 and R2 become closer in the spectrum and are transformed into the high-Q mode (R1) and low-Q mode (R2) due to the interference of the radiative channels between each other. The strong confinement of the electric field inside the structure can greatly strengthen the intensity of the nonlinear signal with the increasing Q-factor of R1, which can reach nearly 2.7×10\textsuperscript4 at $1.7^\circ$ incident angle (yz-plane). We achieve two orders of enhancement of the input electric field amplitude within the slab (as shown in in the insect of Fig. \ref{fig3}(c)). This will significantly enhance the nonlinear generation in the following section.

The two resonance coupled process is indicated by the electric field distribution in xoz plane at the different points of R1 and R2 (shown in Fig. \ref{fig3}(b) and the points are shown in Fig. \ref{fig3}(a)). We further investigate the mechanism of the interaction between the two resonances using the multipolar decomposition and then we explore the transformations of coupled multipoles in the strong coupling process. We perform the Cartesian multipolar expansion to calculate the contributions of toroidal dipole mode \cite{he2018toroidal,gurvitz2019high}. The toroidal dipole mode is a bounded state due to zero overlap between its mode profiles and the waves of the radiative continuum in this subdiffractive periodic system. By careful design, the non-radiating toroidal dipole mode can be excited by coupling from radiative multipoles, such as the electric dipole mode, which offers a new way of constructing the quasi-BIC state. The scattering cross section of the excited multipoles at $1.7^\circ$ incident angle (yz-plane) in Fig. \ref{fig4}(a) indicates the strong magnetic dipole component of R1 and strong electric dipole contribution of R2 \cite{grahn2012electromagnetic}. We further analyse the scattering cross section based on the Cartesian multipole expansion for the case at $\Gamma$ point and under the incident angle of $1.7^\circ$, respectively,  as shown in Fig. \ref{fig4}(b-d). R1 and R2 are mainly governed by the electric toroidal dipole mode (Te) and magnetic dipole in the x-direction (Mx) separately. When the two resonances R1 and R2 are at the narrow spectral gap under the incident angle of $1.7^\circ$, their interference leads to a strong excitation of the electric toroidal moment and boosts its scattering cross section.

\subsection{Second-harmonic generation}

\begin{figure*}
\centering
\includegraphics[width=0.95\textwidth]{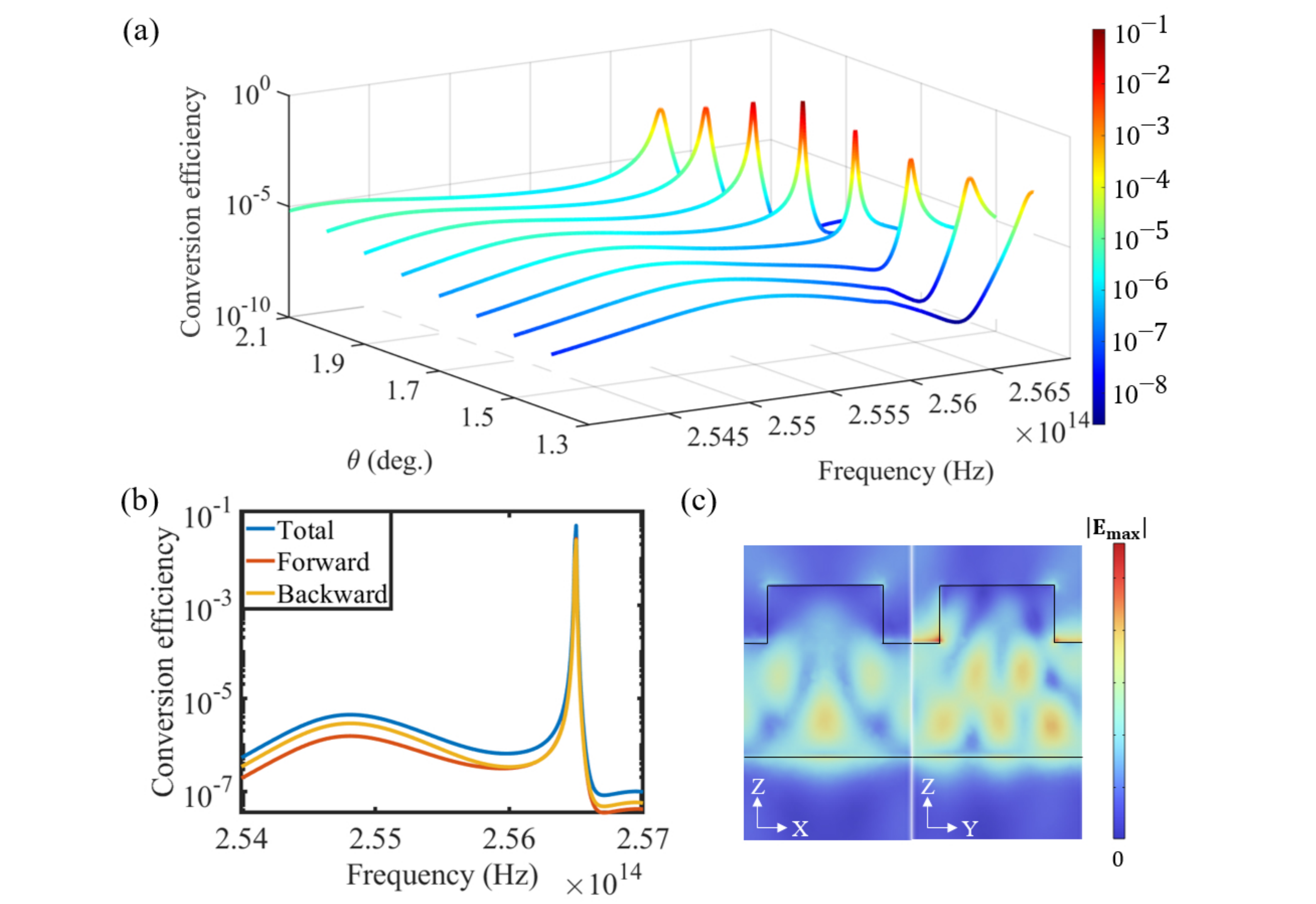}
\caption{\label{fig5}(a) The 3D map of SH emission intensity as a function of frequency and incident angle ($\theta$) along the y-direction. (b) The conversion efficiency under the 1.7-degree incident angle (yz-plane) with the crystal axis shown as Eq. 2. (c) The electric patterns under the fundamental frequency 256.5 THz with the maximum conversion efficiency.}
\end{figure*}

When designing the nanostructure for SHG, it is important to obtain a large mode volume for light-matter interactions in the nonlinear material. In this case, guided modes in the slab provide a perfect platform for achieving large mode volume and enhancing the intensity of the nonlinear signal. This is why we select the nanoparticle array on top instead of adding a hole for each unit to break the continuous translational symmetry. Furthermore, the diagonal second-order nonlinear susceptibility tensor and high transparency \cite{carletti2019second,vakulov2020oxygen} in the visible range of LiNbO3 make it an ideal candidate to apply in the metasurface-slab compared to III-V semiconductors. Besides, the use of LiNbO\textsubscript3 shows that the guided mode in the subwavelength thickness slab can support greatly enhanced light-matter interaction with low-index materials ($n\sim2$).

Next, we simulate the nonlinear effect boosted by the quasi-BIC state based on the relationship between each component of second-order polarisation density $(P_x^{SH},P_y^{SH},P_z^{SH})$ and the electric field of pump light $(E_x,E_y  ,E_z)$ in LiNbO\textsubscript3, which is shown below:
\begin{equation} \label{eq.2}
\begin{bmatrix} P_x^{SH} \\ P_x^{SH} \\ P_x^{SH} \end{bmatrix}
= 2\varepsilon_0 \left[
\begin{array}{cccccc}
     0 & 0 & 0 & 0 &  d_{22} & d_{31} \\ d_{31} & d_{31} & d_{33} & 0 & 0 & 0 \\ d_{22} & 0 & -d_{22} & d_{31} & 0 & 0\\
\end{array}
\right]
\begin{bmatrix} E_x^2 \\ E_y^2 \\ E_z^2 \\ 2E_yE_z \\ 2E_xE_z \\ 2E_xE_y
\end{bmatrix} \, .
\end{equation}
Here, $\varepsilon_0$ is the vacuum permittivity. We set the second-order nonlinear coefficients \cite{zelmon1997infrared} as $d_{31}= \mathrm{-3.2~pm/V}$, $d_{22}=\mathrm{1.9~pm/V}$, $d_{33}=\mathrm{19.5~pm/V}$.

 We define the SHG conversion efficiency $\eta=P_{SHG}/P_{in}$ to describe the ability of a structure to convert the input light into the SH light, where $P_{SHG}$ is the radiated power of SHG within a single unit, and $P_{in}$ is the input power per single unit at the fundamental frequency. The intensity of the generated nonlinear light is highly related to the intensity of the confined electric field within the slab structure. Fig. \ref{fig5}(a) shows the SHG conversion efficiency versus the pump wavelength and incident angle. As the incident angle increases away from the $\Gamma$ point, the SHG conversion efficiency increases. A pump light with an incident angle of $1.7^\circ$ and an input power of 0.4 MW/cm\textsuperscript2 at 256.5 THz ($\sim1170~\mathrm{nm}$) leads to a SH conversion efficiency of around 5\% at 513 THz ($\sim585~\mathrm{nm}$). Such a record-high SHG conversion efficiency from LiNbO\textsubscript3 metasurface consists of 2.6\% for forwarding SH emission, and 2.3\% for the backward SH emission.  This high SHG corresponds to the spectral position of R1 with Q-factor $2.7\times10^4$ as indicated in Fig. \ref{fig3}(b). The Q-factor of R1 at $1.7^\circ$ incident angle is shown in Fig. \ref{fig3}(c) and the corresponding conversion efficiency in Fig. \ref{fig5}(a) reveals that in the metasurface-slab system, the nonlinear response is highly controllable by tuning the characteristics of resonances at the fundamental frequency.
 
 The nonlinear conversion efficiency is not only affected by the resonance at the pump frequency, but also affected by many other factors. Taking SHG as an example, the expression $P_{2\omega}\propto\kappa_2 Q_2 L_2 \kappa_{12} (\kappa_1 Q_1 L_1 P_\omega)^2$ describes the conversion process from the pump light into the SH radiation \cite{koshelev2020subwavelength}, where $P_\omega$ is the input power of the pump light, $Q_1$  and $Q_2$ are the Q-factors of the resonances at the fundamental frequency and the harmonic frequency. $\kappa_1$ and $\kappa_2$ are the coupling factors at the pump and harmonic frequency that the far-field radiation couples to the bound states of the system. $\kappa_{12}$  is the cross-coupling coefficient that depends on the second-order nonlinear susceptibility tensor of the nonlinear material. The expression suggests that three factors can affect the conversion efficiency of SHG: 1) the resonance at the fundamental frequency, 2) the resonance at the SH frequency, and 3) the cross-coupling between the fundamental and SH light. 
 
 \begin{figure*}
\centering
\includegraphics[width=0.8\textwidth]{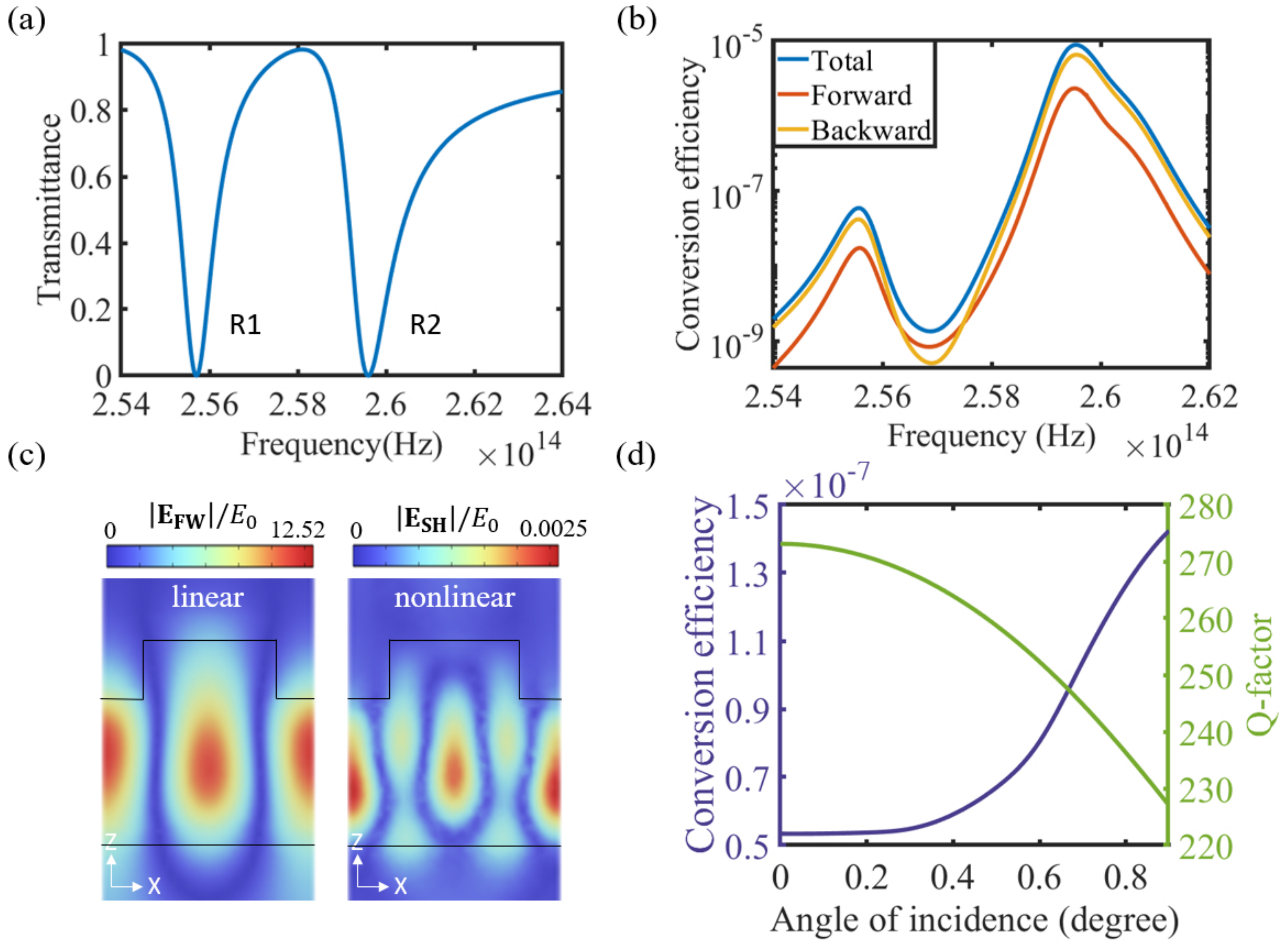}
\caption{\label{fig6}(a) The transmittance spectrum of the metasurface-slab system at the gamma point. (b) The simulated SH conversion efficiency versus the frequency at the gamma point with different refractive index at harmonic frequency. (c) The electric field profiles of R1 at the fundamental wave (FW) frequency and SH frequency. (d) The conversion efficiency and Q-factor of R1 versus incident (yz-plane).}
\end{figure*}
 
At the $\Gamma$ point, the BIC state at the harmonic frequency has no coupling to the far-field radiative channel ($Q_2\longrightarrow~\infty, \kappa_2\longrightarrow~\infty, Q_2\kappa_2\longrightarrow~0$). This will lead to an enhanced SH field that does not radiate to the far-field domain, which is indicated as resonance-forbidden SHG \cite{jin2021resonance}. Here we further demonstrate an analogous phenomenon: anomalous SHG in the metasurface-slab system (shown in Fig. \ref{fig6}). Fig. \ref{fig6}(a) shows that R1 and R2 at $\Gamma$ point with finite Q-factor exhibit the odd distribution under $C_2^z$ in the structure. The calculated field distribution of SHG (Fig. \ref{fig6}(c)) is even under $C_2^z$. This electric field is further decoupled to the far-field radiative channel with the odd distribution, which results in the SHG signal propagating and oscillating through the whole infinite metasurfaces system along the x-axis and y-axis with a weak leakage to the far-field in the z-direction. In most cases, the higher Q-factor and stronger confinement of the electric field would enhance the nonlinear generations. However, via tuning the incident angle (yz-plane), we observe the anomalous SHG in our designed structure, showing that the conversion efficiency increases with the decrease of the Q-factor of R1 (see Fig. \ref{fig6}(d)). 

The protection of $C_2^z$ symmetry is broken by tuning the incident angle, which results in a stronger leakage of the electromagnetic energy to the far-field at the harmonic frequency. For further demonstration, we calculate the electric patterns in different incident angles and their maximum intensity of electric fields (see Supplementary Material S6). When we increase the incident angle (yz-plane), the maximum intensity of the electric field at harmonic frequency increases while the maximum intensity at fundamental frequency decreases. The process of the tuning of conversion efficiency near the $\Gamma$ point demonstrates this unique effect of BICs at harmonic frequency on the SHG emission. In addition, the coupling of SH mode and far-field radiation, and the influence on the conversion process are other interesting topics worth further discussion.

\begin{figure*}
\centering
\includegraphics[width=1.0\textwidth]{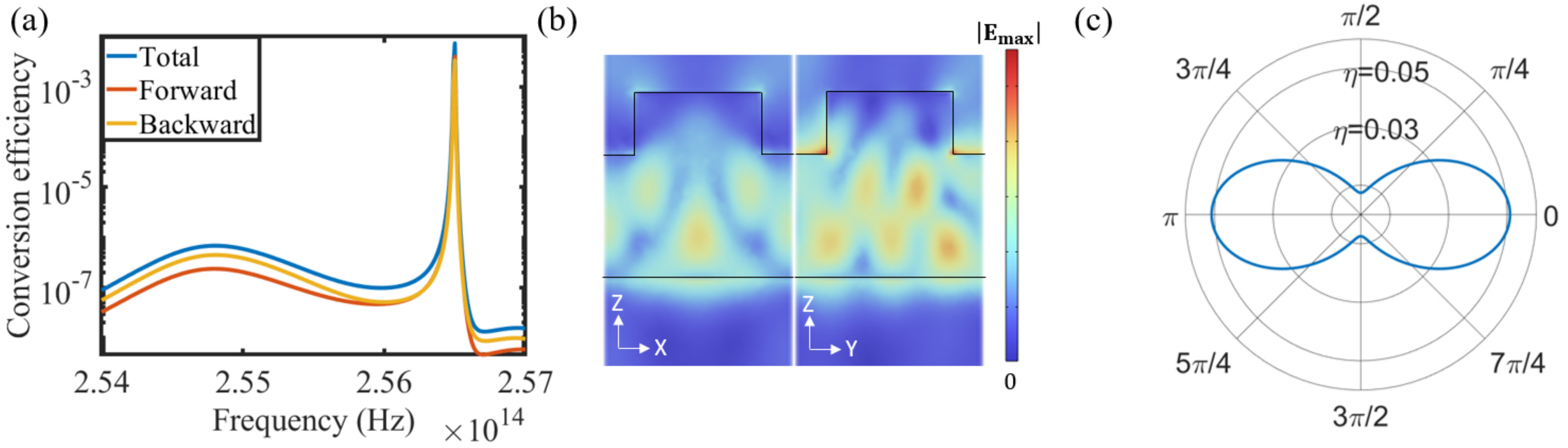}
\caption{\label{fig7}(a) The conversion efficiency under 1.7-degree incident angle (yz-plane) with the crystal axis shown as Eq. \ref{eq.3}. (b) The electric field profiles under the fundamental frequency 256.5 THz with the maximum conversion efficiency. (c) The conversion efficiency tuning with the rotation of the crystal axis along the y-axis.}
\end{figure*}

The SHG is also closely related to the cross-coupling between the fundamental and SH light. The cross coupling, $\kappa_{12}$, can be tuned by controlling the second-order nonlinear susceptibility tensor of the nonlinear material. Here, we rotate the crystal axis for $90^\circ$ along the y-axis (parallel to the extraordinary axis) and change the tensor to
\begin{equation}\label{eq.3}
\begin{bmatrix} P_x^{SH} \\ P_x^{SH} \\ P_x^{SH} \end{bmatrix}=2\varepsilon_0\begin{bmatrix} -d_{22} & 0 & d_{22} & 0 & 0 & d_{31} \\ d_{33} & d_{31} & d_{31} & 0 & 0 & 0 \\ 0 & 0 & 0 & d_{31} & d_{22} & 0 \end{bmatrix} \begin{bmatrix} E_x^2 \\ E_y^2 \\ E_z^2 \\ 2E_yE_z \\ 2E_xE_z \\ 2E_xE_y \end{bmatrix}\, .
\end{equation}
The second-order nonlinear coefficients are the same as we set in Eq.~\ref{eq.2}. Based on Eq.~\ref{eq.3}, we calculate the spectrum of SHG conversion efficiency at an incident angle of $1.7^\circ$ and an input power of 0.4 MW/cm\textsuperscript2, as shown in Fig. \ref{fig7}(a). Fig. \ref{fig7}(b) shows the field distribution at the peak of the SHG efficiency, which has the same spectral position as R2. The total efficiency decreases to only 0.7\% with 0.4\% for the forward radiation and 0.3\% for the backward radiation. We keep the refractive index of LiNbO\textsubscript3 unchanged for the extraordinary axis, and in parallel to the y-axis, so that R1 remains the same, and so does the optical response in the linear region. 
The active tuning of conversion efficiency can be observed in Fig. \ref{fig7}(c) by continuously rotating the crystal axis along the y-axis. 
We note that the difference between the conversion efficiency in Fig. \ref{fig5}(b) and Fig. \ref{fig7}(a) is induced only by the changes of $\kappa_{12}$, indicating a strong effect of cross-coupling between the fundamental and SH light.

\section{Conclusion}
In summary, we have demonstrated the potential of a new platform for enhanced SHG in low-refractive-index materials. By utilising a nanoparticles-on-top metasurfaces, we successfully transform the inaccessible odd guided modes into QBIC-type resonances by engineering the symmetry of the infinite slab. By tuning the incident pump angle, we further generated a high-Q resonance with suppressed radiation channels based on the strong coupling region between the two designed resonances. This high-Q resonance originates from the Friedrich-Wintgen BIC, and its Q-factor can be controlled by adjusting the spectral gap between two resonances. The conversion efficiency is also greatly boosted by the strong confinement of the electric field within the LiNbO\textsubscript3 flim at the resonant position. Meanwhile, by observing the anomalous SHG and changing the crystal axis, we also provided evidence of the influence of resonance at emission wavelength and cross-coupling between fundamental and SH light. This approach of constructing the strong light-matter interaction to generate high conversion efficiency can be genetically applied to different nonlinear materials including both low and high refractive indices, as well as different nonlinear processes in addition to the SHG. Our results offer unique opportunities for the design and application of all-dielectric resonance-enhanced harmonic generations.

\begin{acknowledgments}
Z.Z. acknowledges the support from the Royal Society scholarship. L.X. and M.R. acknowledge support from the UK Research and Innovation Future Leaders Fellowship (MR/T040513/1). L.X, C.Y. and M.R. acknowledge the support from the Quality Research (QR) funding at NTU. D.A.S. acknowledges financial support from the ARC Discovery Early Career Researcher Award (DE19010043). The authors acknowledge the use of NTU High Performance Computing cluster Hamilton. 
\end{acknowledgments}



\begin{thebibliography}{49}%
\makeatletter
\providecommand \@ifxundefined [1]{%
 \@ifx{#1\undefined}
}%
\providecommand \@ifnum [1]{%
 \ifnum #1\expandafter \@firstoftwo
 \else \expandafter \@secondoftwo
 \fi
}%
\providecommand \@ifx [1]{%
 \ifx #1\expandafter \@firstoftwo
 \else \expandafter \@secondoftwo
 \fi
}%
\providecommand \natexlab [1]{#1}%
\providecommand \enquote  [1]{``#1''}%
\providecommand \bibnamefont  [1]{#1}%
\providecommand \bibfnamefont [1]{#1}%
\providecommand \citenamefont [1]{#1}%
\providecommand \href@noop [0]{\@secondoftwo}%
\providecommand \href [0]{\begingroup \@sanitize@url \@href}%
\providecommand \@href[1]{\@@startlink{#1}\@@href}%
\providecommand \@@href[1]{\endgroup#1\@@endlink}%
\providecommand \@sanitize@url [0]{\catcode `\\12\catcode `\$12\catcode
  `\&12\catcode `\#12\catcode `\^12\catcode `\_12\catcode `\%12\relax}%
\providecommand \@@startlink[1]{}%
\providecommand \@@endlink[0]{}%
\providecommand \url  [0]{\begingroup\@sanitize@url \@url }%
\providecommand \@url [1]{\endgroup\@href {#1}{\urlprefix }}%
\providecommand \urlprefix  [0]{URL }%
\providecommand \Eprint [0]{\href }%
\providecommand \doibase [0]{http://dx.doi.org/}%
\providecommand \selectlanguage [0]{\@gobble}%
\providecommand \bibinfo  [0]{\@secondoftwo}%
\providecommand \bibfield  [0]{\@secondoftwo}%
\providecommand \translation [1]{[#1]}%
\providecommand \BibitemOpen [0]{}%
\providecommand \bibitemStop [0]{}%
\providecommand \bibitemNoStop [0]{.\EOS\space}%
\providecommand \EOS [0]{\spacefactor3000\relax}%
\providecommand \BibitemShut  [1]{\csname bibitem#1\endcsname}%
\let\auto@bib@innerbib\@empty
\bibitem [{\citenamefont {Boyd}(2020)}]{boyd2020nonlinear}%
  \BibitemOpen
  \bibfield  {author} {\bibinfo {author} {\bibfnamefont {R.~W.}\ \bibnamefont
  {Boyd}},\ }\href@noop {} {\emph {\bibinfo {title} {Nonlinear optics}}}\
  (\bibinfo  {publisher} {Academic press},\ \bibinfo {year} {2020})\BibitemShut
  {NoStop}%
\bibitem [{\citenamefont {Mesch}\ \emph {et~al.}(2016)\citenamefont {Mesch},
  \citenamefont {Metzger}, \citenamefont {Hentschel},\ and\ \citenamefont
  {Giessen}}]{mesch2016nonlinear}%
  \BibitemOpen
  \bibfield  {author} {\bibinfo {author} {\bibfnamefont {M.}~\bibnamefont
  {Mesch}}, \bibinfo {author} {\bibfnamefont {B.}~\bibnamefont {Metzger}},
  \bibinfo {author} {\bibfnamefont {M.}~\bibnamefont {Hentschel}}, \ and\
  \bibinfo {author} {\bibfnamefont {H.}~\bibnamefont {Giessen}},\ }\href@noop
  {} {\bibfield  {journal} {\bibinfo  {journal} {Nano Letters}\ }\textbf
  {\bibinfo {volume} {16}},\ \bibinfo {pages} {3155} (\bibinfo {year}
  {2016})}\BibitemShut {NoStop}%
\bibitem [{\citenamefont {Butet}\ \emph {et~al.}(2012)\citenamefont {Butet},
  \citenamefont {Russier-Antoine}, \citenamefont {Jonin}, \citenamefont
  {Lascoux}, \citenamefont {Benichou},\ and\ \citenamefont
  {Brevet}}]{butet2012sensing}%
  \BibitemOpen
  \bibfield  {author} {\bibinfo {author} {\bibfnamefont {J.}~\bibnamefont
  {Butet}}, \bibinfo {author} {\bibfnamefont {I.}~\bibnamefont
  {Russier-Antoine}}, \bibinfo {author} {\bibfnamefont {C.}~\bibnamefont
  {Jonin}}, \bibinfo {author} {\bibfnamefont {N.}~\bibnamefont {Lascoux}},
  \bibinfo {author} {\bibfnamefont {E.}~\bibnamefont {Benichou}}, \ and\
  \bibinfo {author} {\bibfnamefont {P.-F.}\ \bibnamefont {Brevet}},\
  }\href@noop {} {\bibfield  {journal} {\bibinfo  {journal} {Nano Letters}\
  }\textbf {\bibinfo {volume} {12}},\ \bibinfo {pages} {1697} (\bibinfo {year}
  {2012})}\BibitemShut {NoStop}%
\bibitem [{\citenamefont {Moody}\ \emph {et~al.}(2020)\citenamefont {Moody},
  \citenamefont {Chang}, \citenamefont {Steiner},\ and\ \citenamefont
  {Bowers}}]{moody2020chip}%
  \BibitemOpen
  \bibfield  {author} {\bibinfo {author} {\bibfnamefont {G.}~\bibnamefont
  {Moody}}, \bibinfo {author} {\bibfnamefont {L.}~\bibnamefont {Chang}},
  \bibinfo {author} {\bibfnamefont {T.~J.}\ \bibnamefont {Steiner}}, \ and\
  \bibinfo {author} {\bibfnamefont {J.~E.}\ \bibnamefont {Bowers}},\
  }\href@noop {} {\bibfield  {journal} {\bibinfo  {journal} {AVS Quantum
  Science}\ }\textbf {\bibinfo {volume} {2}},\ \bibinfo {pages} {041702}
  (\bibinfo {year} {2020})}\BibitemShut {NoStop}%
\bibitem [{\citenamefont {Schlickriede}\ \emph {et~al.}(2020)\citenamefont
  {Schlickriede}, \citenamefont {Kruk}, \citenamefont {Wang}, \citenamefont
  {Sain}, \citenamefont {Kivshar},\ and\ \citenamefont
  {Zentgraf}}]{schlickriede2020nonlinear}%
  \BibitemOpen
  \bibfield  {author} {\bibinfo {author} {\bibfnamefont {C.}~\bibnamefont
  {Schlickriede}}, \bibinfo {author} {\bibfnamefont {S.~S.}\ \bibnamefont
  {Kruk}}, \bibinfo {author} {\bibfnamefont {L.}~\bibnamefont {Wang}}, \bibinfo
  {author} {\bibfnamefont {B.}~\bibnamefont {Sain}}, \bibinfo {author}
  {\bibfnamefont {Y.}~\bibnamefont {Kivshar}}, \ and\ \bibinfo {author}
  {\bibfnamefont {T.}~\bibnamefont {Zentgraf}},\ }\href@noop {} {\bibfield
  {journal} {\bibinfo  {journal} {Nano Letters}\ }\textbf {\bibinfo {volume}
  {20}},\ \bibinfo {pages} {4370} (\bibinfo {year} {2020})}\BibitemShut
  {NoStop}%
\bibitem [{\citenamefont {Burns}\ \emph {et~al.}(2000)\citenamefont {Burns},
  \citenamefont {Simpson},\ and\ \citenamefont
  {Averkiou}}]{burns2000nonlinear}%
  \BibitemOpen
  \bibfield  {author} {\bibinfo {author} {\bibfnamefont {P.~N.}\ \bibnamefont
  {Burns}}, \bibinfo {author} {\bibfnamefont {D.~H.}\ \bibnamefont {Simpson}},
  \ and\ \bibinfo {author} {\bibfnamefont {M.~A.}\ \bibnamefont {Averkiou}},\
  }\href@noop {} {\bibfield  {journal} {\bibinfo  {journal} {Ultrasound in
  Medicine and Biology}\ }\textbf {\bibinfo {volume} {26}},\ \bibinfo {pages}
  {S19} (\bibinfo {year} {2000})}\BibitemShut {NoStop}%
\bibitem [{\citenamefont {Vabishchevich}\ \emph {et~al.}(2018)\citenamefont
  {Vabishchevich}, \citenamefont {Liu}, \citenamefont {Sinclair}, \citenamefont
  {Keeler}, \citenamefont {Peake},\ and\ \citenamefont
  {Brener}}]{vabishchevich2018enhanced}%
  \BibitemOpen
  \bibfield  {author} {\bibinfo {author} {\bibfnamefont {P.~P.}\ \bibnamefont
  {Vabishchevich}}, \bibinfo {author} {\bibfnamefont {S.}~\bibnamefont {Liu}},
  \bibinfo {author} {\bibfnamefont {M.~B.}\ \bibnamefont {Sinclair}}, \bibinfo
  {author} {\bibfnamefont {G.~A.}\ \bibnamefont {Keeler}}, \bibinfo {author}
  {\bibfnamefont {G.~M.}\ \bibnamefont {Peake}}, \ and\ \bibinfo {author}
  {\bibfnamefont {I.}~\bibnamefont {Brener}},\ }\href@noop {} {\bibfield
  {journal} {\bibinfo  {journal} {ACS Photonics}\ }\textbf {\bibinfo {volume}
  {5}},\ \bibinfo {pages} {1685} (\bibinfo {year} {2018})}\BibitemShut
  {NoStop}%
\bibitem [{\citenamefont {Kruk}\ \emph {et~al.}(2015)\citenamefont {Kruk},
  \citenamefont {Weismann}, \citenamefont {Bykov}, \citenamefont {Mamonov},
  \citenamefont {Kolmychek}, \citenamefont {Murzina}, \citenamefont {Panoiu},
  \citenamefont {Neshev},\ and\ \citenamefont {Kivshar}}]{kruk2015enhanced}%
  \BibitemOpen
  \bibfield  {author} {\bibinfo {author} {\bibfnamefont {S.}~\bibnamefont
  {Kruk}}, \bibinfo {author} {\bibfnamefont {M.}~\bibnamefont {Weismann}},
  \bibinfo {author} {\bibfnamefont {A.~Y.}\ \bibnamefont {Bykov}}, \bibinfo
  {author} {\bibfnamefont {E.~A.}\ \bibnamefont {Mamonov}}, \bibinfo {author}
  {\bibfnamefont {I.~A.}\ \bibnamefont {Kolmychek}}, \bibinfo {author}
  {\bibfnamefont {T.}~\bibnamefont {Murzina}}, \bibinfo {author} {\bibfnamefont
  {N.~C.}\ \bibnamefont {Panoiu}}, \bibinfo {author} {\bibfnamefont {D.~N.}\
  \bibnamefont {Neshev}}, \ and\ \bibinfo {author} {\bibfnamefont {Y.~S.}\
  \bibnamefont {Kivshar}},\ }\href@noop {} {\bibfield  {journal} {\bibinfo
  {journal} {ACS Photonics}\ }\textbf {\bibinfo {volume} {2}},\ \bibinfo
  {pages} {1007} (\bibinfo {year} {2015})}\BibitemShut {NoStop}%
\bibitem [{\citenamefont {Zheludev}\ and\ \citenamefont
  {Kivshar}(2012)}]{zheludev2012metamaterials}%
  \BibitemOpen
  \bibfield  {author} {\bibinfo {author} {\bibfnamefont {N.~I.}\ \bibnamefont
  {Zheludev}}\ and\ \bibinfo {author} {\bibfnamefont {Y.~S.}\ \bibnamefont
  {Kivshar}},\ }\href@noop {} {\bibfield  {journal} {\bibinfo  {journal}
  {Nature Materials}\ }\textbf {\bibinfo {volume} {11}},\ \bibinfo {pages}
  {917} (\bibinfo {year} {2012})}\BibitemShut {NoStop}%
\bibitem [{\citenamefont {Kauranen}\ and\ \citenamefont
  {Zayats}(2012)}]{kauranen2012nonlinear}%
  \BibitemOpen
  \bibfield  {author} {\bibinfo {author} {\bibfnamefont {M.}~\bibnamefont
  {Kauranen}}\ and\ \bibinfo {author} {\bibfnamefont {A.~V.}\ \bibnamefont
  {Zayats}},\ }\href@noop {} {\bibfield  {journal} {\bibinfo  {journal} {Nature
  Photonics}\ }\textbf {\bibinfo {volume} {6}},\ \bibinfo {pages} {737}
  (\bibinfo {year} {2012})}\BibitemShut {NoStop}%
\bibitem [{\citenamefont {Rahmani}\ \emph {et~al.}(2018)\citenamefont
  {Rahmani}, \citenamefont {Leo}, \citenamefont {Brener}, \citenamefont
  {Zayats}, \citenamefont {Maier}, \citenamefont {De~Angelis}, \citenamefont
  {Tan}, \citenamefont {Gili}, \citenamefont {Karouta}, \citenamefont {Oulton}
  \emph {et~al.}}]{rahmani2018nonlinear}%
  \BibitemOpen
  \bibfield  {author} {\bibinfo {author} {\bibfnamefont {M.}~\bibnamefont
  {Rahmani}}, \bibinfo {author} {\bibfnamefont {G.}~\bibnamefont {Leo}},
  \bibinfo {author} {\bibfnamefont {I.}~\bibnamefont {Brener}}, \bibinfo
  {author} {\bibfnamefont {A.~V.}\ \bibnamefont {Zayats}}, \bibinfo {author}
  {\bibfnamefont {S.~A.}\ \bibnamefont {Maier}}, \bibinfo {author}
  {\bibfnamefont {C.}~\bibnamefont {De~Angelis}}, \bibinfo {author}
  {\bibfnamefont {H.}~\bibnamefont {Tan}}, \bibinfo {author} {\bibfnamefont
  {V.~F.}\ \bibnamefont {Gili}}, \bibinfo {author} {\bibfnamefont
  {F.}~\bibnamefont {Karouta}}, \bibinfo {author} {\bibfnamefont
  {R.}~\bibnamefont {Oulton}},  \emph {et~al.},\ }\href@noop {} {\bibfield
  {journal} {\bibinfo  {journal} {Opto-Electronic Advances}\ }\textbf {\bibinfo
  {volume} {1}},\ \bibinfo {pages} {180021} (\bibinfo {year}
  {2018})}\BibitemShut {NoStop}%
\bibitem [{\citenamefont {Carletti}\ \emph
  {et~al.}(2019{\natexlab{a}})\citenamefont {Carletti}, \citenamefont {Kruk},
  \citenamefont {Bogdanov}, \citenamefont {De~Angelis},\ and\ \citenamefont
  {Kivshar}}]{carletti2019high}%
  \BibitemOpen
  \bibfield  {author} {\bibinfo {author} {\bibfnamefont {L.}~\bibnamefont
  {Carletti}}, \bibinfo {author} {\bibfnamefont {S.~S.}\ \bibnamefont {Kruk}},
  \bibinfo {author} {\bibfnamefont {A.~A.}\ \bibnamefont {Bogdanov}}, \bibinfo
  {author} {\bibfnamefont {C.}~\bibnamefont {De~Angelis}}, \ and\ \bibinfo
  {author} {\bibfnamefont {Y.}~\bibnamefont {Kivshar}},\ }\href@noop {}
  {\bibfield  {journal} {\bibinfo  {journal} {Physical Review Research}\
  }\textbf {\bibinfo {volume} {1}},\ \bibinfo {pages} {023016} (\bibinfo {year}
  {2019}{\natexlab{a}})}\BibitemShut {NoStop}%
\bibitem [{\citenamefont {Kuznetsov}\ \emph {et~al.}(2016)\citenamefont
  {Kuznetsov}, \citenamefont {Miroshnichenko}, \citenamefont {Brongersma},
  \citenamefont {Kivshar},\ and\ \citenamefont
  {Lukâ€™yanchuk}}]{kuznetsov2016optically}%
  \BibitemOpen
  \bibfield  {author} {\bibinfo {author} {\bibfnamefont {A.~I.}\ \bibnamefont
  {Kuznetsov}}, \bibinfo {author} {\bibfnamefont {A.~E.}\ \bibnamefont
  {Miroshnichenko}}, \bibinfo {author} {\bibfnamefont {M.~L.}\ \bibnamefont
  {Brongersma}}, \bibinfo {author} {\bibfnamefont {Y.~S.}\ \bibnamefont
  {Kivshar}}, \ and\ \bibinfo {author} {\bibfnamefont {B.}~\bibnamefont
  {Lukâ€™yanchuk}},\ }\href@noop {} {\bibfield  {journal} {\bibinfo  {journal}
  {Science}\ }\textbf {\bibinfo {volume} {354}},\ \bibinfo {pages} {aag2472}
  (\bibinfo {year} {2016})}\BibitemShut {NoStop}%
\bibitem [{\citenamefont {Bonacina}\ \emph {et~al.}(2020)\citenamefont
  {Bonacina}, \citenamefont {Brevet}, \citenamefont {Finazzi},\ and\
  \citenamefont {Celebrano}}]{bonacina2020harmonic}%
  \BibitemOpen
  \bibfield  {author} {\bibinfo {author} {\bibfnamefont {L.}~\bibnamefont
  {Bonacina}}, \bibinfo {author} {\bibfnamefont {P.-F.}\ \bibnamefont
  {Brevet}}, \bibinfo {author} {\bibfnamefont {M.}~\bibnamefont {Finazzi}}, \
  and\ \bibinfo {author} {\bibfnamefont {M.}~\bibnamefont {Celebrano}},\
  }\href@noop {} {\bibfield  {journal} {\bibinfo  {journal} {Journal of Applied
  Physics}\ }\textbf {\bibinfo {volume} {127}},\ \bibinfo {pages} {230901}
  (\bibinfo {year} {2020})}\BibitemShut {NoStop}%
\bibitem [{\citenamefont {Carletti}\ \emph {et~al.}(2015)\citenamefont
  {Carletti}, \citenamefont {Locatelli}, \citenamefont {Stepanenko},
  \citenamefont {Leo},\ and\ \citenamefont
  {De~Angelis}}]{carletti2015enhanced}%
  \BibitemOpen
  \bibfield  {author} {\bibinfo {author} {\bibfnamefont {L.}~\bibnamefont
  {Carletti}}, \bibinfo {author} {\bibfnamefont {A.}~\bibnamefont {Locatelli}},
  \bibinfo {author} {\bibfnamefont {O.}~\bibnamefont {Stepanenko}}, \bibinfo
  {author} {\bibfnamefont {G.}~\bibnamefont {Leo}}, \ and\ \bibinfo {author}
  {\bibfnamefont {C.}~\bibnamefont {De~Angelis}},\ }\href@noop {} {\bibfield
  {journal} {\bibinfo  {journal} {Optics Express}\ }\textbf {\bibinfo {volume}
  {23}},\ \bibinfo {pages} {26544} (\bibinfo {year} {2015})}\BibitemShut
  {NoStop}%
\bibitem [{\citenamefont {Grinblat}(2021)}]{grinblat2021nonlinear}%
  \BibitemOpen
  \bibfield  {author} {\bibinfo {author} {\bibfnamefont {G.}~\bibnamefont
  {Grinblat}},\ }\href@noop {} {\bibfield  {journal} {\bibinfo  {journal} {ACS
  Photonics}\ }\textbf {\bibinfo {volume} {8}},\ \bibinfo {pages} {3406}
  (\bibinfo {year} {2021})}\BibitemShut {NoStop}%
\bibitem [{\citenamefont {Shcherbakov}\ \emph {et~al.}(2014)\citenamefont
  {Shcherbakov}, \citenamefont {Neshev}, \citenamefont {Hopkins}, \citenamefont
  {Shorokhov}, \citenamefont {Staude}, \citenamefont {Melik-Gaykazyan},
  \citenamefont {Decker}, \citenamefont {Ezhov}, \citenamefont
  {Miroshnichenko}, \citenamefont {Brener} \emph
  {et~al.}}]{shcherbakov2014enhanced}%
  \BibitemOpen
  \bibfield  {author} {\bibinfo {author} {\bibfnamefont {M.~R.}\ \bibnamefont
  {Shcherbakov}}, \bibinfo {author} {\bibfnamefont {D.~N.}\ \bibnamefont
  {Neshev}}, \bibinfo {author} {\bibfnamefont {B.}~\bibnamefont {Hopkins}},
  \bibinfo {author} {\bibfnamefont {A.~S.}\ \bibnamefont {Shorokhov}}, \bibinfo
  {author} {\bibfnamefont {I.}~\bibnamefont {Staude}}, \bibinfo {author}
  {\bibfnamefont {E.~V.}\ \bibnamefont {Melik-Gaykazyan}}, \bibinfo {author}
  {\bibfnamefont {M.}~\bibnamefont {Decker}}, \bibinfo {author} {\bibfnamefont
  {A.~A.}\ \bibnamefont {Ezhov}}, \bibinfo {author} {\bibfnamefont {A.~E.}\
  \bibnamefont {Miroshnichenko}}, \bibinfo {author} {\bibfnamefont
  {I.}~\bibnamefont {Brener}},  \emph {et~al.},\ }\href@noop {} {\bibfield
  {journal} {\bibinfo  {journal} {Nano Letters}\ }\textbf {\bibinfo {volume}
  {14}},\ \bibinfo {pages} {6488} (\bibinfo {year} {2014})}\BibitemShut
  {NoStop}%
\bibitem [{\citenamefont {Grinblat}\ \emph {et~al.}(2016)\citenamefont
  {Grinblat}, \citenamefont {Li}, \citenamefont {Nielsen}, \citenamefont
  {Oulton},\ and\ \citenamefont {Maier}}]{grinblat2016enhanced}%
  \BibitemOpen
  \bibfield  {author} {\bibinfo {author} {\bibfnamefont {G.}~\bibnamefont
  {Grinblat}}, \bibinfo {author} {\bibfnamefont {Y.}~\bibnamefont {Li}},
  \bibinfo {author} {\bibfnamefont {M.~P.}\ \bibnamefont {Nielsen}}, \bibinfo
  {author} {\bibfnamefont {R.~F.}\ \bibnamefont {Oulton}}, \ and\ \bibinfo
  {author} {\bibfnamefont {S.~A.}\ \bibnamefont {Maier}},\ }\href@noop {}
  {\bibfield  {journal} {\bibinfo  {journal} {Nano Letters}\ }\textbf {\bibinfo
  {volume} {16}},\ \bibinfo {pages} {4635} (\bibinfo {year}
  {2016})}\BibitemShut {NoStop}%
\bibitem [{\citenamefont {LoÌˆchner}\ \emph {et~al.}(2018)\citenamefont
  {LoÌˆchner}, \citenamefont {Fedotova}, \citenamefont {Liu}, \citenamefont
  {Keeler}, \citenamefont {Peake}, \citenamefont {Saravi}, \citenamefont
  {Shcherbakov}, \citenamefont {Burger}, \citenamefont {Fedyanin},
  \citenamefont {Brener} \emph {et~al.}}]{locher2018polarization}%
  \BibitemOpen
  \bibfield  {author} {\bibinfo {author} {\bibfnamefont {F.~J.}\ \bibnamefont
  {LoÌˆchner}}, \bibinfo {author} {\bibfnamefont {A.~N.}\ \bibnamefont
  {Fedotova}}, \bibinfo {author} {\bibfnamefont {S.}~\bibnamefont {Liu}},
  \bibinfo {author} {\bibfnamefont {G.~A.}\ \bibnamefont {Keeler}}, \bibinfo
  {author} {\bibfnamefont {G.~M.}\ \bibnamefont {Peake}}, \bibinfo {author}
  {\bibfnamefont {S.}~\bibnamefont {Saravi}}, \bibinfo {author} {\bibfnamefont
  {M.~R.}\ \bibnamefont {Shcherbakov}}, \bibinfo {author} {\bibfnamefont
  {S.}~\bibnamefont {Burger}}, \bibinfo {author} {\bibfnamefont {A.~A.}\
  \bibnamefont {Fedyanin}}, \bibinfo {author} {\bibfnamefont {I.}~\bibnamefont
  {Brener}},  \emph {et~al.},\ }\href@noop {} {\bibfield  {journal} {\bibinfo
  {journal} {Acs Photonics}\ }\textbf {\bibinfo {volume} {5}},\ \bibinfo
  {pages} {1786} (\bibinfo {year} {2018})}\BibitemShut {NoStop}%
\bibitem [{\citenamefont {Liu}\ \emph {et~al.}(2016)\citenamefont {Liu},
  \citenamefont {Sinclair}, \citenamefont {Saravi}, \citenamefont {Keeler},
  \citenamefont {Yang}, \citenamefont {Reno}, \citenamefont {Peake},
  \citenamefont {Setzpfandt}, \citenamefont {Staude}, \citenamefont {Pertsch}
  \emph {et~al.}}]{liu2016resonantly}%
  \BibitemOpen
  \bibfield  {author} {\bibinfo {author} {\bibfnamefont {S.}~\bibnamefont
  {Liu}}, \bibinfo {author} {\bibfnamefont {M.~B.}\ \bibnamefont {Sinclair}},
  \bibinfo {author} {\bibfnamefont {S.}~\bibnamefont {Saravi}}, \bibinfo
  {author} {\bibfnamefont {G.~A.}\ \bibnamefont {Keeler}}, \bibinfo {author}
  {\bibfnamefont {Y.}~\bibnamefont {Yang}}, \bibinfo {author} {\bibfnamefont
  {J.}~\bibnamefont {Reno}}, \bibinfo {author} {\bibfnamefont {G.~M.}\
  \bibnamefont {Peake}}, \bibinfo {author} {\bibfnamefont {F.}~\bibnamefont
  {Setzpfandt}}, \bibinfo {author} {\bibfnamefont {I.}~\bibnamefont {Staude}},
  \bibinfo {author} {\bibfnamefont {T.}~\bibnamefont {Pertsch}},  \emph
  {et~al.},\ }\href@noop {} {\bibfield  {journal} {\bibinfo  {journal} {Nano
  letters}\ }\textbf {\bibinfo {volume} {16}},\ \bibinfo {pages} {5426}
  (\bibinfo {year} {2016})}\BibitemShut {NoStop}%
\bibitem [{\citenamefont {Lu}\ \emph {et~al.}(2022)\citenamefont {Lu},
  \citenamefont {Xu}, \citenamefont {Ouyang}, \citenamefont {Xian},
  \citenamefont {Cao}, \citenamefont {Chen},\ and\ \citenamefont
  {Li}}]{YudongLu2022Opto}%
  \BibitemOpen
  \bibfield  {author} {\bibinfo {author} {\bibfnamefont {Y.}~\bibnamefont
  {Lu}}, \bibinfo {author} {\bibfnamefont {Y.}~\bibnamefont {Xu}}, \bibinfo
  {author} {\bibfnamefont {X.}~\bibnamefont {Ouyang}}, \bibinfo {author}
  {\bibfnamefont {M.}~\bibnamefont {Xian}}, \bibinfo {author} {\bibfnamefont
  {Y.}~\bibnamefont {Cao}}, \bibinfo {author} {\bibfnamefont {K.}~\bibnamefont
  {Chen}}, \ and\ \bibinfo {author} {\bibfnamefont {X.}~\bibnamefont {Li}},\
  }\href@noop {} {\enquote {\bibinfo {title} {Cylindrical vector beams reveal
  radiationless anapole condition in a resonant state},}\ } (\bibinfo {year}
  {2022})\BibitemShut {NoStop}%
\bibitem [{\citenamefont {Gigli}\ and\ \citenamefont
  {Leo}(2022)}]{gigli2022all}%
  \BibitemOpen
  \bibfield  {author} {\bibinfo {author} {\bibfnamefont {C.}~\bibnamefont
  {Gigli}}\ and\ \bibinfo {author} {\bibfnamefont {G.}~\bibnamefont {Leo}},\
  }\href@noop {} {\bibfield  {journal} {\bibinfo  {journal} {Opto-Electronic
  Advances}\ ,\ \bibinfo {pages} {210093}} (\bibinfo {year}
  {2022})}\BibitemShut {NoStop}%
\bibitem [{\citenamefont {Melik-Gaykazyan}\ \emph {et~al.}(2018)\citenamefont
  {Melik-Gaykazyan}, \citenamefont {Kruk}, \citenamefont {Camacho-Morales},
  \citenamefont {Xu}, \citenamefont {Rahmani}, \citenamefont {Zangeneh~Kamali},
  \citenamefont {Lamprianidis}, \citenamefont {Miroshnichenko}, \citenamefont
  {Fedyanin}, \citenamefont {Neshev},\ and\ \citenamefont
  {Kivshar}}]{doi:10.1021/acsphotonics.7b01277}%
  \BibitemOpen
  \bibfield  {author} {\bibinfo {author} {\bibfnamefont {E.~V.}\ \bibnamefont
  {Melik-Gaykazyan}}, \bibinfo {author} {\bibfnamefont {S.~S.}\ \bibnamefont
  {Kruk}}, \bibinfo {author} {\bibfnamefont {R.}~\bibnamefont
  {Camacho-Morales}}, \bibinfo {author} {\bibfnamefont {L.}~\bibnamefont {Xu}},
  \bibinfo {author} {\bibfnamefont {M.}~\bibnamefont {Rahmani}}, \bibinfo
  {author} {\bibfnamefont {K.}~\bibnamefont {Zangeneh~Kamali}}, \bibinfo
  {author} {\bibfnamefont {A.}~\bibnamefont {Lamprianidis}}, \bibinfo {author}
  {\bibfnamefont {A.~E.}\ \bibnamefont {Miroshnichenko}}, \bibinfo {author}
  {\bibfnamefont {A.~A.}\ \bibnamefont {Fedyanin}}, \bibinfo {author}
  {\bibfnamefont {D.~N.}\ \bibnamefont {Neshev}}, \ and\ \bibinfo {author}
  {\bibfnamefont {Y.~S.}\ \bibnamefont {Kivshar}},\ }\href@noop {} {\bibfield
  {journal} {\bibinfo  {journal} {ACS Photonics}\ }\textbf {\bibinfo {volume}
  {5}},\ \bibinfo {pages} {728} (\bibinfo {year} {2018})}\BibitemShut {NoStop}%
\bibitem [{\citenamefont {Frizyuk}\ \emph {et~al.}(2019)\citenamefont
  {Frizyuk}, \citenamefont {Volkovskaya}, \citenamefont {Smirnova},
  \citenamefont {Poddubny},\ and\ \citenamefont {Petrov}}]{frizyuk2019second}%
  \BibitemOpen
  \bibfield  {author} {\bibinfo {author} {\bibfnamefont {K.}~\bibnamefont
  {Frizyuk}}, \bibinfo {author} {\bibfnamefont {I.}~\bibnamefont
  {Volkovskaya}}, \bibinfo {author} {\bibfnamefont {D.}~\bibnamefont
  {Smirnova}}, \bibinfo {author} {\bibfnamefont {A.}~\bibnamefont {Poddubny}},
  \ and\ \bibinfo {author} {\bibfnamefont {M.}~\bibnamefont {Petrov}},\
  }\href@noop {} {\bibfield  {journal} {\bibinfo  {journal} {Physical Review
  B}\ }\textbf {\bibinfo {volume} {99}},\ \bibinfo {pages} {075425} (\bibinfo
  {year} {2019})}\BibitemShut {NoStop}%
\bibitem [{\citenamefont {Liang}\ \emph {et~al.}(2020)\citenamefont {Liang},
  \citenamefont {Koshelev}, \citenamefont {Zhang}, \citenamefont {Lin},
  \citenamefont {Lin}, \citenamefont {Wu}, \citenamefont {Jia},\ and\
  \citenamefont {Kivshar}}]{liang2020bound}%
  \BibitemOpen
  \bibfield  {author} {\bibinfo {author} {\bibfnamefont {Y.}~\bibnamefont
  {Liang}}, \bibinfo {author} {\bibfnamefont {K.}~\bibnamefont {Koshelev}},
  \bibinfo {author} {\bibfnamefont {F.}~\bibnamefont {Zhang}}, \bibinfo
  {author} {\bibfnamefont {H.}~\bibnamefont {Lin}}, \bibinfo {author}
  {\bibfnamefont {S.}~\bibnamefont {Lin}}, \bibinfo {author} {\bibfnamefont
  {J.}~\bibnamefont {Wu}}, \bibinfo {author} {\bibfnamefont {B.}~\bibnamefont
  {Jia}}, \ and\ \bibinfo {author} {\bibfnamefont {Y.}~\bibnamefont
  {Kivshar}},\ }\href@noop {} {\bibfield  {journal} {\bibinfo  {journal} {Nano
  Letters}\ }\textbf {\bibinfo {volume} {20}},\ \bibinfo {pages} {6351}
  (\bibinfo {year} {2020})}\BibitemShut {NoStop}%
\bibitem [{\citenamefont {Zeng}\ \emph {et~al.}(2021)\citenamefont {Zeng},
  \citenamefont {Liu}, \citenamefont {Wang},\ and\ \citenamefont
  {Lin}}]{zeng2021light}%
  \BibitemOpen
  \bibfield  {author} {\bibinfo {author} {\bibfnamefont {T.-Y.}\ \bibnamefont
  {Zeng}}, \bibinfo {author} {\bibfnamefont {G.-D.}\ \bibnamefont {Liu}},
  \bibinfo {author} {\bibfnamefont {L.-L.}\ \bibnamefont {Wang}}, \ and\
  \bibinfo {author} {\bibfnamefont {Q.}~\bibnamefont {Lin}},\ }\href@noop {}
  {\bibfield  {journal} {\bibinfo  {journal} {Optics Express}\ }\textbf
  {\bibinfo {volume} {29}},\ \bibinfo {pages} {40177} (\bibinfo {year}
  {2021})}\BibitemShut {NoStop}%
\bibitem [{\citenamefont {Koshelev}\ \emph {et~al.}(2020)\citenamefont
  {Koshelev}, \citenamefont {Kruk}, \citenamefont {Melik-Gaykazyan},
  \citenamefont {Choi}, \citenamefont {Bogdanov}, \citenamefont {Park},\ and\
  \citenamefont {Kivshar}}]{koshelev2020subwavelength}%
  \BibitemOpen
  \bibfield  {author} {\bibinfo {author} {\bibfnamefont {K.}~\bibnamefont
  {Koshelev}}, \bibinfo {author} {\bibfnamefont {S.}~\bibnamefont {Kruk}},
  \bibinfo {author} {\bibfnamefont {E.}~\bibnamefont {Melik-Gaykazyan}},
  \bibinfo {author} {\bibfnamefont {J.-H.}\ \bibnamefont {Choi}}, \bibinfo
  {author} {\bibfnamefont {A.}~\bibnamefont {Bogdanov}}, \bibinfo {author}
  {\bibfnamefont {H.-G.}\ \bibnamefont {Park}}, \ and\ \bibinfo {author}
  {\bibfnamefont {Y.}~\bibnamefont {Kivshar}},\ }\href@noop {} {\bibfield
  {journal} {\bibinfo  {journal} {Science}\ }\textbf {\bibinfo {volume}
  {367}},\ \bibinfo {pages} {288} (\bibinfo {year} {2020})}\BibitemShut
  {NoStop}%
\bibitem [{\citenamefont {Yang}\ \emph {et~al.}(2020)\citenamefont {Yang},
  \citenamefont {Liu}, \citenamefont {Gan}, \citenamefont {Fang}, \citenamefont
  {Han},\ and\ \citenamefont {Hao}}]{yang2020nonlinear}%
  \BibitemOpen
  \bibfield  {author} {\bibinfo {author} {\bibfnamefont {Q.}~\bibnamefont
  {Yang}}, \bibinfo {author} {\bibfnamefont {Y.}~\bibnamefont {Liu}}, \bibinfo
  {author} {\bibfnamefont {X.}~\bibnamefont {Gan}}, \bibinfo {author}
  {\bibfnamefont {C.}~\bibnamefont {Fang}}, \bibinfo {author} {\bibfnamefont
  {G.}~\bibnamefont {Han}}, \ and\ \bibinfo {author} {\bibfnamefont
  {Y.}~\bibnamefont {Hao}},\ }\href@noop {} {\bibfield  {journal} {\bibinfo
  {journal} {IEEE Photonics Journal}\ }\textbf {\bibinfo {volume} {12}},\
  \bibinfo {pages} {1} (\bibinfo {year} {2020})}\BibitemShut {NoStop}%
\bibitem [{\citenamefont {Tong}\ \emph {et~al.}(2016)\citenamefont {Tong},
  \citenamefont {Gong}, \citenamefont {Liu}, \citenamefont {Yuan},
  \citenamefont {Huang}, \citenamefont {Xia},\ and\ \citenamefont
  {Wang}}]{tong2016enhanced}%
  \BibitemOpen
  \bibfield  {author} {\bibinfo {author} {\bibfnamefont {W.}~\bibnamefont
  {Tong}}, \bibinfo {author} {\bibfnamefont {C.}~\bibnamefont {Gong}}, \bibinfo
  {author} {\bibfnamefont {X.}~\bibnamefont {Liu}}, \bibinfo {author}
  {\bibfnamefont {S.}~\bibnamefont {Yuan}}, \bibinfo {author} {\bibfnamefont
  {Q.}~\bibnamefont {Huang}}, \bibinfo {author} {\bibfnamefont
  {J.}~\bibnamefont {Xia}}, \ and\ \bibinfo {author} {\bibfnamefont
  {Y.}~\bibnamefont {Wang}},\ }\href@noop {} {\bibfield  {journal} {\bibinfo
  {journal} {Optics Express}\ }\textbf {\bibinfo {volume} {24}},\ \bibinfo
  {pages} {19661} (\bibinfo {year} {2016})}\BibitemShut {NoStop}%
\bibitem [{\citenamefont {Anthur}\ \emph {et~al.}(2020)\citenamefont {Anthur},
  \citenamefont {Zhang}, \citenamefont {Paniagua-Dominguez}, \citenamefont
  {Kalashnikov}, \citenamefont {Ha}, \citenamefont {Ma{\ss}}, \citenamefont
  {Kuznetsov},\ and\ \citenamefont {Krivitsky}}]{anthur2020continuous}%
  \BibitemOpen
  \bibfield  {author} {\bibinfo {author} {\bibfnamefont {A.~P.}\ \bibnamefont
  {Anthur}}, \bibinfo {author} {\bibfnamefont {H.}~\bibnamefont {Zhang}},
  \bibinfo {author} {\bibfnamefont {R.}~\bibnamefont {Paniagua-Dominguez}},
  \bibinfo {author} {\bibfnamefont {D.~A.}\ \bibnamefont {Kalashnikov}},
  \bibinfo {author} {\bibfnamefont {S.~T.}\ \bibnamefont {Ha}}, \bibinfo
  {author} {\bibfnamefont {T.~W.}\ \bibnamefont {Ma{\ss}}}, \bibinfo {author}
  {\bibfnamefont {A.~I.}\ \bibnamefont {Kuznetsov}}, \ and\ \bibinfo {author}
  {\bibfnamefont {L.}~\bibnamefont {Krivitsky}},\ }\href@noop {} {\bibfield
  {journal} {\bibinfo  {journal} {Nano Letters}\ }\textbf {\bibinfo {volume}
  {20}},\ \bibinfo {pages} {8745} (\bibinfo {year} {2020})}\BibitemShut
  {NoStop}%
\bibitem [{\citenamefont {Koshelev}\ \emph {et~al.}(2019)\citenamefont
  {Koshelev}, \citenamefont {Tang}, \citenamefont {Li}, \citenamefont {Choi},
  \citenamefont {Li},\ and\ \citenamefont {Kivshar}}]{koshelev2019nonlinear}%
  \BibitemOpen
  \bibfield  {author} {\bibinfo {author} {\bibfnamefont {K.}~\bibnamefont
  {Koshelev}}, \bibinfo {author} {\bibfnamefont {Y.}~\bibnamefont {Tang}},
  \bibinfo {author} {\bibfnamefont {K.}~\bibnamefont {Li}}, \bibinfo {author}
  {\bibfnamefont {D.-Y.}\ \bibnamefont {Choi}}, \bibinfo {author}
  {\bibfnamefont {G.}~\bibnamefont {Li}}, \ and\ \bibinfo {author}
  {\bibfnamefont {Y.}~\bibnamefont {Kivshar}},\ }\href@noop {} {\bibfield
  {journal} {\bibinfo  {journal} {ACS Photonics}\ }\textbf {\bibinfo {volume}
  {6}},\ \bibinfo {pages} {1639} (\bibinfo {year} {2019})}\BibitemShut
  {NoStop}%
\bibitem [{\citenamefont {Xu}\ \emph {et~al.}(2019)\citenamefont {Xu},
  \citenamefont {Zangeneh~Kamali}, \citenamefont {Huang}, \citenamefont
  {Rahmani}, \citenamefont {Smirnov}, \citenamefont {Camacho-Morales},
  \citenamefont {Ma}, \citenamefont {Zhang}, \citenamefont {Woolley},
  \citenamefont {Neshev} \emph {et~al.}}]{xu2019dynamic}%
  \BibitemOpen
  \bibfield  {author} {\bibinfo {author} {\bibfnamefont {L.}~\bibnamefont
  {Xu}}, \bibinfo {author} {\bibfnamefont {K.}~\bibnamefont {Zangeneh~Kamali}},
  \bibinfo {author} {\bibfnamefont {L.}~\bibnamefont {Huang}}, \bibinfo
  {author} {\bibfnamefont {M.}~\bibnamefont {Rahmani}}, \bibinfo {author}
  {\bibfnamefont {A.}~\bibnamefont {Smirnov}}, \bibinfo {author} {\bibfnamefont
  {R.}~\bibnamefont {Camacho-Morales}}, \bibinfo {author} {\bibfnamefont
  {Y.}~\bibnamefont {Ma}}, \bibinfo {author} {\bibfnamefont {G.}~\bibnamefont
  {Zhang}}, \bibinfo {author} {\bibfnamefont {M.}~\bibnamefont {Woolley}},
  \bibinfo {author} {\bibfnamefont {D.}~\bibnamefont {Neshev}},  \emph
  {et~al.},\ }\href@noop {} {\bibfield  {journal} {\bibinfo  {journal}
  {Advanced Science}\ }\textbf {\bibinfo {volume} {6}},\ \bibinfo {pages}
  {1802119} (\bibinfo {year} {2019})}\BibitemShut {NoStop}%
\bibitem [{\citenamefont {Fujii}\ \emph {et~al.}(1977)\citenamefont {Fujii},
  \citenamefont {Yoshida}, \citenamefont {Misawa}, \citenamefont {Maekawa},\
  and\ \citenamefont {Sakudo}}]{fujii1977nonlinear}%
  \BibitemOpen
  \bibfield  {author} {\bibinfo {author} {\bibfnamefont {Y.}~\bibnamefont
  {Fujii}}, \bibinfo {author} {\bibfnamefont {S.}~\bibnamefont {Yoshida}},
  \bibinfo {author} {\bibfnamefont {S.}~\bibnamefont {Misawa}}, \bibinfo
  {author} {\bibfnamefont {S.}~\bibnamefont {Maekawa}}, \ and\ \bibinfo
  {author} {\bibfnamefont {T.}~\bibnamefont {Sakudo}},\ }\href@noop {}
  {\bibfield  {journal} {\bibinfo  {journal} {Applied Physics Letters}\
  }\textbf {\bibinfo {volume} {31}},\ \bibinfo {pages} {815} (\bibinfo {year}
  {1977})}\BibitemShut {NoStop}%
\bibitem [{\citenamefont {Allakhverdiev}\ \emph {et~al.}(2009)\citenamefont
  {Allakhverdiev}, \citenamefont {Yetis}, \citenamefont {{\"O}zbek},
  \citenamefont {Baykara},\ and\ \citenamefont
  {Salaev}}]{allakhverdiev2009effective}%
  \BibitemOpen
  \bibfield  {author} {\bibinfo {author} {\bibfnamefont {K.}~\bibnamefont
  {Allakhverdiev}}, \bibinfo {author} {\bibfnamefont {M.}~\bibnamefont
  {Yetis}}, \bibinfo {author} {\bibfnamefont {S.}~\bibnamefont {{\"O}zbek}},
  \bibinfo {author} {\bibfnamefont {T.}~\bibnamefont {Baykara}}, \ and\
  \bibinfo {author} {\bibfnamefont {E.~Y.}\ \bibnamefont {Salaev}},\
  }\href@noop {} {\bibfield  {journal} {\bibinfo  {journal} {Laser Physics}\
  }\textbf {\bibinfo {volume} {19}},\ \bibinfo {pages} {1092} (\bibinfo {year}
  {2009})}\BibitemShut {NoStop}%
\bibitem [{\citenamefont {Jiang}\ \emph {et~al.}(2020)\citenamefont {Jiang},
  \citenamefont {Hao}, \citenamefont {Ji}, \citenamefont {Hou}, \citenamefont
  {Yi}, \citenamefont {Mao}, \citenamefont {Gan},\ and\ \citenamefont
  {Zhao}}]{jiang2020high}%
  \BibitemOpen
  \bibfield  {author} {\bibinfo {author} {\bibfnamefont {B.}~\bibnamefont
  {Jiang}}, \bibinfo {author} {\bibfnamefont {Z.}~\bibnamefont {Hao}}, \bibinfo
  {author} {\bibfnamefont {Y.}~\bibnamefont {Ji}}, \bibinfo {author}
  {\bibfnamefont {Y.}~\bibnamefont {Hou}}, \bibinfo {author} {\bibfnamefont
  {R.}~\bibnamefont {Yi}}, \bibinfo {author} {\bibfnamefont {D.}~\bibnamefont
  {Mao}}, \bibinfo {author} {\bibfnamefont {X.}~\bibnamefont {Gan}}, \ and\
  \bibinfo {author} {\bibfnamefont {J.}~\bibnamefont {Zhao}},\ }\href@noop {}
  {\bibfield  {journal} {\bibinfo  {journal} {Light: Science \& Applications}\
  }\textbf {\bibinfo {volume} {9}},\ \bibinfo {pages} {1} (\bibinfo {year}
  {2020})}\BibitemShut {NoStop}%
\bibitem [{\citenamefont {Carletti}\ \emph
  {et~al.}(2019{\natexlab{b}})\citenamefont {Carletti}, \citenamefont {Li},
  \citenamefont {Sautter}, \citenamefont {Staude}, \citenamefont {De~Angelis},
  \citenamefont {Li},\ and\ \citenamefont {Neshev}}]{carletti2019second}%
  \BibitemOpen
  \bibfield  {author} {\bibinfo {author} {\bibfnamefont {L.}~\bibnamefont
  {Carletti}}, \bibinfo {author} {\bibfnamefont {C.}~\bibnamefont {Li}},
  \bibinfo {author} {\bibfnamefont {J.}~\bibnamefont {Sautter}}, \bibinfo
  {author} {\bibfnamefont {I.}~\bibnamefont {Staude}}, \bibinfo {author}
  {\bibfnamefont {C.}~\bibnamefont {De~Angelis}}, \bibinfo {author}
  {\bibfnamefont {T.}~\bibnamefont {Li}}, \ and\ \bibinfo {author}
  {\bibfnamefont {D.~N.}\ \bibnamefont {Neshev}},\ }\href@noop {} {\bibfield
  {journal} {\bibinfo  {journal} {Optics Express}\ }\textbf {\bibinfo {volume}
  {27}},\ \bibinfo {pages} {33391} (\bibinfo {year}
  {2019}{\natexlab{b}})}\BibitemShut {NoStop}%
\bibitem [{\citenamefont {Vakulov}\ \emph {et~al.}(2020)\citenamefont
  {Vakulov}, \citenamefont {Zamburg}, \citenamefont {Khakhulin}, \citenamefont
  {Geldash}, \citenamefont {Golosov}, \citenamefont {Zavadski}, \citenamefont
  {Miakonkikh}, \citenamefont {Rudenko}, \citenamefont {Dostanko},
  \citenamefont {He} \emph {et~al.}}]{vakulov2020oxygen}%
  \BibitemOpen
  \bibfield  {author} {\bibinfo {author} {\bibfnamefont {Z.}~\bibnamefont
  {Vakulov}}, \bibinfo {author} {\bibfnamefont {E.}~\bibnamefont {Zamburg}},
  \bibinfo {author} {\bibfnamefont {D.}~\bibnamefont {Khakhulin}}, \bibinfo
  {author} {\bibfnamefont {A.}~\bibnamefont {Geldash}}, \bibinfo {author}
  {\bibfnamefont {D.~A.}\ \bibnamefont {Golosov}}, \bibinfo {author}
  {\bibfnamefont {S.~M.}\ \bibnamefont {Zavadski}}, \bibinfo {author}
  {\bibfnamefont {A.~V.}\ \bibnamefont {Miakonkikh}}, \bibinfo {author}
  {\bibfnamefont {K.~V.}\ \bibnamefont {Rudenko}}, \bibinfo {author}
  {\bibfnamefont {A.~P.}\ \bibnamefont {Dostanko}}, \bibinfo {author}
  {\bibfnamefont {Z.}~\bibnamefont {He}},  \emph {et~al.},\ }\href@noop {}
  {\bibfield  {journal} {\bibinfo  {journal} {Nanomaterials}\ }\textbf
  {\bibinfo {volume} {10}},\ \bibinfo {pages} {1371} (\bibinfo {year}
  {2020})}\BibitemShut {NoStop}%
\bibitem [{\citenamefont {Gao}\ \emph {et~al.}(2016)\citenamefont {Gao},
  \citenamefont {Hsu}, \citenamefont {Zhen}, \citenamefont {Lin}, \citenamefont
  {Joannopoulos}, \citenamefont {Solja{\v{c}}i{\'c}},\ and\ \citenamefont
  {Chen}}]{gao2016formation}%
  \BibitemOpen
  \bibfield  {author} {\bibinfo {author} {\bibfnamefont {X.}~\bibnamefont
  {Gao}}, \bibinfo {author} {\bibfnamefont {C.~W.}\ \bibnamefont {Hsu}},
  \bibinfo {author} {\bibfnamefont {B.}~\bibnamefont {Zhen}}, \bibinfo {author}
  {\bibfnamefont {X.}~\bibnamefont {Lin}}, \bibinfo {author} {\bibfnamefont
  {J.~D.}\ \bibnamefont {Joannopoulos}}, \bibinfo {author} {\bibfnamefont
  {M.}~\bibnamefont {Solja{\v{c}}i{\'c}}}, \ and\ \bibinfo {author}
  {\bibfnamefont {H.}~\bibnamefont {Chen}},\ }\href@noop {} {\bibfield
  {journal} {\bibinfo  {journal} {Scientific Reports}\ }\textbf {\bibinfo
  {volume} {6}},\ \bibinfo {pages} {1} (\bibinfo {year} {2016})}\BibitemShut
  {NoStop}%
\bibitem [{\citenamefont {Yin}\ \emph {et~al.}(2020)\citenamefont {Yin},
  \citenamefont {Jin}, \citenamefont {Solja{\v{c}}i{\'c}}, \citenamefont
  {Peng},\ and\ \citenamefont {Zhen}}]{yin2020observation}%
  \BibitemOpen
  \bibfield  {author} {\bibinfo {author} {\bibfnamefont {X.}~\bibnamefont
  {Yin}}, \bibinfo {author} {\bibfnamefont {J.}~\bibnamefont {Jin}}, \bibinfo
  {author} {\bibfnamefont {M.}~\bibnamefont {Solja{\v{c}}i{\'c}}}, \bibinfo
  {author} {\bibfnamefont {C.}~\bibnamefont {Peng}}, \ and\ \bibinfo {author}
  {\bibfnamefont {B.}~\bibnamefont {Zhen}},\ }\href@noop {} {\bibfield
  {journal} {\bibinfo  {journal} {Nature}\ }\textbf {\bibinfo {volume} {580}},\
  \bibinfo {pages} {467} (\bibinfo {year} {2020})}\BibitemShut {NoStop}%
\bibitem [{\citenamefont {Hsu}\ \emph {et~al.}(2013)\citenamefont {Hsu},
  \citenamefont {Zhen}, \citenamefont {Lee}, \citenamefont {Chua},
  \citenamefont {Johnson}, \citenamefont {Joannopoulos},\ and\ \citenamefont
  {Solja{\v{c}}i{\'c}}}]{hsu2013observation}%
  \BibitemOpen
  \bibfield  {author} {\bibinfo {author} {\bibfnamefont {C.~W.}\ \bibnamefont
  {Hsu}}, \bibinfo {author} {\bibfnamefont {B.}~\bibnamefont {Zhen}}, \bibinfo
  {author} {\bibfnamefont {J.}~\bibnamefont {Lee}}, \bibinfo {author}
  {\bibfnamefont {S.-L.}\ \bibnamefont {Chua}}, \bibinfo {author}
  {\bibfnamefont {S.~G.}\ \bibnamefont {Johnson}}, \bibinfo {author}
  {\bibfnamefont {J.~D.}\ \bibnamefont {Joannopoulos}}, \ and\ \bibinfo
  {author} {\bibfnamefont {M.}~\bibnamefont {Solja{\v{c}}i{\'c}}},\ }\href@noop
  {} {\bibfield  {journal} {\bibinfo  {journal} {Nature}\ }\textbf {\bibinfo
  {volume} {499}},\ \bibinfo {pages} {188} (\bibinfo {year}
  {2013})}\BibitemShut {NoStop}%
\bibitem [{\citenamefont {Zelmon}\ \emph {et~al.}(1997)\citenamefont {Zelmon},
  \citenamefont {Small},\ and\ \citenamefont {Jundt}}]{zelmon1997infrared}%
  \BibitemOpen
  \bibfield  {author} {\bibinfo {author} {\bibfnamefont {D.~E.}\ \bibnamefont
  {Zelmon}}, \bibinfo {author} {\bibfnamefont {D.~L.}\ \bibnamefont {Small}}, \
  and\ \bibinfo {author} {\bibfnamefont {D.}~\bibnamefont {Jundt}},\
  }\href@noop {} {\bibfield  {journal} {\bibinfo  {journal} {JOSA B}\ }\textbf
  {\bibinfo {volume} {14}},\ \bibinfo {pages} {3319} (\bibinfo {year}
  {1997})}\BibitemShut {NoStop}%
\bibitem [{\citenamefont {Joannopoulos}\ \emph {et~al.}(2011)\citenamefont
  {Joannopoulos}, \citenamefont {Johnson}, \citenamefont {Winn},\ and\
  \citenamefont {Meade}}]{joannopoulos2011photonic}%
  \BibitemOpen
  \bibfield  {author} {\bibinfo {author} {\bibfnamefont {J.~D.}\ \bibnamefont
  {Joannopoulos}}, \bibinfo {author} {\bibfnamefont {S.~G.}\ \bibnamefont
  {Johnson}}, \bibinfo {author} {\bibfnamefont {J.~N.}\ \bibnamefont {Winn}}, \
  and\ \bibinfo {author} {\bibfnamefont {R.~D.}\ \bibnamefont {Meade}},\ }in\
  \href@noop {} {\emph {\bibinfo {booktitle} {Photonic Crystals}}}\ (\bibinfo
  {publisher} {Princeton university press},\ \bibinfo {year}
  {2011})\BibitemShut {NoStop}%
\bibitem [{\citenamefont {Huang}\ \emph {et~al.}(2013)\citenamefont {Huang},
  \citenamefont {Yu},\ and\ \citenamefont {Cao}}]{huang2013general}%
  \BibitemOpen
  \bibfield  {author} {\bibinfo {author} {\bibfnamefont {L.}~\bibnamefont
  {Huang}}, \bibinfo {author} {\bibfnamefont {Y.}~\bibnamefont {Yu}}, \ and\
  \bibinfo {author} {\bibfnamefont {L.}~\bibnamefont {Cao}},\ }\href@noop {}
  {\bibfield  {journal} {\bibinfo  {journal} {Nano Letters}\ }\textbf {\bibinfo
  {volume} {13}},\ \bibinfo {pages} {3559} (\bibinfo {year}
  {2013})}\BibitemShut {NoStop}%
\bibitem [{\citenamefont {Huang}\ \emph {et~al.}(2021)\citenamefont {Huang},
  \citenamefont {Xu}, \citenamefont {Rahmani}, \citenamefont {Neshev},\ and\
  \citenamefont {Miroshnichenko}}]{huang2021pushing}%
  \BibitemOpen
  \bibfield  {author} {\bibinfo {author} {\bibfnamefont {L.}~\bibnamefont
  {Huang}}, \bibinfo {author} {\bibfnamefont {L.}~\bibnamefont {Xu}}, \bibinfo
  {author} {\bibfnamefont {M.}~\bibnamefont {Rahmani}}, \bibinfo {author}
  {\bibfnamefont {D.}~\bibnamefont {Neshev}}, \ and\ \bibinfo {author}
  {\bibfnamefont {A.~E.}\ \bibnamefont {Miroshnichenko}},\ }\href@noop {}
  {\bibfield  {journal} {\bibinfo  {journal} {Advanced Photonics}\ }\textbf
  {\bibinfo {volume} {3}},\ \bibinfo {pages} {016004} (\bibinfo {year}
  {2021})}\BibitemShut {NoStop}%
\bibitem [{\citenamefont {Wiersig}(2006)}]{wiersig2006formation}%
  \BibitemOpen
  \bibfield  {author} {\bibinfo {author} {\bibfnamefont {J.}~\bibnamefont
  {Wiersig}},\ }\href@noop {} {\bibfield  {journal} {\bibinfo  {journal}
  {Physical Review Letters}\ }\textbf {\bibinfo {volume} {97}},\ \bibinfo
  {pages} {253901} (\bibinfo {year} {2006})}\BibitemShut {NoStop}%
\bibitem [{\citenamefont {He}\ \emph {et~al.}(2018)\citenamefont {He},
  \citenamefont {Guo}, \citenamefont {Feng}, \citenamefont {Xu},\ and\
  \citenamefont {Miroshnichenko}}]{he2018toroidal}%
  \BibitemOpen
  \bibfield  {author} {\bibinfo {author} {\bibfnamefont {Y.}~\bibnamefont
  {He}}, \bibinfo {author} {\bibfnamefont {G.}~\bibnamefont {Guo}}, \bibinfo
  {author} {\bibfnamefont {T.}~\bibnamefont {Feng}}, \bibinfo {author}
  {\bibfnamefont {Y.}~\bibnamefont {Xu}}, \ and\ \bibinfo {author}
  {\bibfnamefont {A.~E.}\ \bibnamefont {Miroshnichenko}},\ }\href@noop {}
  {\bibfield  {journal} {\bibinfo  {journal} {Physical Review B}\ }\textbf
  {\bibinfo {volume} {98}},\ \bibinfo {pages} {161112} (\bibinfo {year}
  {2018})}\BibitemShut {NoStop}%
\bibitem [{\citenamefont {Gurvitz}\ \emph {et~al.}(2019)\citenamefont
  {Gurvitz}, \citenamefont {Ladutenko}, \citenamefont {Dergachev},
  \citenamefont {Evlyukhin}, \citenamefont {Miroshnichenko},\ and\
  \citenamefont {Shalin}}]{gurvitz2019high}%
  \BibitemOpen
  \bibfield  {author} {\bibinfo {author} {\bibfnamefont {E.~A.}\ \bibnamefont
  {Gurvitz}}, \bibinfo {author} {\bibfnamefont {K.~S.}\ \bibnamefont
  {Ladutenko}}, \bibinfo {author} {\bibfnamefont {P.~A.}\ \bibnamefont
  {Dergachev}}, \bibinfo {author} {\bibfnamefont {A.~B.}\ \bibnamefont
  {Evlyukhin}}, \bibinfo {author} {\bibfnamefont {A.~E.}\ \bibnamefont
  {Miroshnichenko}}, \ and\ \bibinfo {author} {\bibfnamefont {A.~S.}\
  \bibnamefont {Shalin}},\ }\href@noop {} {\bibfield  {journal} {\bibinfo
  {journal} {Laser \& Photonics Reviews}\ }\textbf {\bibinfo {volume} {13}},\
  \bibinfo {pages} {1800266} (\bibinfo {year} {2019})}\BibitemShut {NoStop}%
\bibitem [{\citenamefont {Grahn}\ \emph {et~al.}(2012)\citenamefont {Grahn},
  \citenamefont {Shevchenko},\ and\ \citenamefont
  {Kaivola}}]{grahn2012electromagnetic}%
  \BibitemOpen
  \bibfield  {author} {\bibinfo {author} {\bibfnamefont {P.}~\bibnamefont
  {Grahn}}, \bibinfo {author} {\bibfnamefont {A.}~\bibnamefont {Shevchenko}}, \
  and\ \bibinfo {author} {\bibfnamefont {M.}~\bibnamefont {Kaivola}},\
  }\href@noop {} {\bibfield  {journal} {\bibinfo  {journal} {New Journal of
  Physics}\ }\textbf {\bibinfo {volume} {14}},\ \bibinfo {pages} {093033}
  (\bibinfo {year} {2012})}\BibitemShut {NoStop}%
\bibitem [{\citenamefont {Jin}\ \emph {et~al.}(2021)\citenamefont {Jin},
  \citenamefont {Lu},\ and\ \citenamefont {Zhen}}]{jin2021resonance}%
  \BibitemOpen
  \bibfield  {author} {\bibinfo {author} {\bibfnamefont {J.}~\bibnamefont
  {Jin}}, \bibinfo {author} {\bibfnamefont {J.}~\bibnamefont {Lu}}, \ and\
  \bibinfo {author} {\bibfnamefont {B.}~\bibnamefont {Zhen}},\ }\href@noop {}
  {\bibfield  {journal} {\bibinfo  {journal} {Nanophotonics}\ }\textbf
  {\bibinfo {volume} {10}},\ \bibinfo {pages} {4233} (\bibinfo {year}
  {2021})}\BibitemShut {NoStop}%
\end{thebibliography}

\providecommand{\noopsort}[1]{}\providecommand{\singleletter}[1]{#1}%

\end{document}